\documentclass[twocolumn,journal,10pt,letterpaper]{IEEEtran}
\usepackage[pdftex]{graphicx,color}
\usepackage[dvipsnames]{xcolor}
\usepackage{amsmath,amsbsy,amsfonts,amssymb}
\usepackage{mathrsfs}
\usepackage{mathtools}
\usepackage{cases}
\mathtoolsset{showonlyrefs}
\ifCLASSOPTIONcompsoc
\usepackage[caption=false,font=normalsize,labelfont=sf,textfont=sf]{subfig}
\else
\usepackage[caption=false,font=footnotesize]{subfig}
\fi
\usepackage{dblfloatfix}
\usepackage{flushend}
\usepackage[bookmarks=false]{hyperref}

\DeclarePairedDelimiter\floor{\lfloor}{\rfloor}

\definecolor{IndianRed3}{rgb}{0.8,0.33,0.33}

\newcommand\Tstrut{\rule{0pt}{2.6ex}}         
\newcommand\Bstrut{\rule[-0.9ex]{0pt}{0pt}}   

\usepackage{acro}
\DeclareAcronym{AWGN}{short = AWGN ,long = additive white gaussian noise}
\DeclareAcronym{AoI}{short = AoI ,long = age of information}
\DeclareAcronym{CDF}{short = CDF ,long = cumulative distribution function}
\DeclareAcronym{CRA}{short = CRA ,long = contention resolution ALOHA}
\DeclareAcronym{CRDSA}{short = CRDSA ,long = contention resolution diversity slotted ALOHA}
\DeclareAcronym{CSA}{short = CSA ,long = coded slotted ALOHA}
\DeclareAcronym{C-RAN}{short = C-RAN ,long = cloud radio access network}
\DeclareAcronym{DAMA}{short = DAMA ,long = demand assigned multiple access}
\DeclareAcronym{DSA}{short = DSA ,long = diversity slotted ALOHA}
\DeclareAcronym{eMBB}{short = eMBB ,long = enhanced mobile broadband}
\DeclareAcronym{FEC}{short = FEC ,long = forward error correction}
\DeclareAcronym{GEO}{short = GEO ,long = geostationary orbit}
\DeclareAcronym{HAP}{short = HAP ,long = high-altitude platform,foreign-plural={}}
\DeclareAcronym{IC}{short = IC ,long = interference cancellation}
\DeclareAcronym{IoT}{short = IoT ,long = internet of things}
\DeclareAcronym{IRSA}{short = IRSA ,long = irregular repetition slotted ALOHA}
\DeclareAcronym{LEO}{short = LEO ,long = low Earth orbit}
\DeclareAcronym{M2M}{short = M2M ,long = machine-to-machine}
\DeclareAcronym{MAC}{short = MAC ,long = medium access}
\DeclareAcronym{MPR}{short = MPR ,long = multi-packet reception}
\DeclareAcronym{MTC}{short = MTC ,long = machine-type communications}
\DeclareAcronym{mMTC}{short = mMTC ,long = massive machine-type communications}
\DeclareAcronym{NTN}{short = NTN ,long = non-terrestrial network,foreign-plural = {}}
\DeclareAcronym{PDF}{short = PDF ,long = probability density function}
\DeclareAcronym{PER}{short = PER ,long = packet error rate}
\DeclareAcronym{PLR}{short = PLR ,long = packet loss rate}
\DeclareAcronym{PMF}{short = PMF ,long = probability mass function}
\DeclareAcronym{RA}{short = RA ,long = random access}
\DeclareAcronym{RRH}{short = RRH ,long = remote radio head,foreign-plural = {}}
\DeclareAcronym{SA}{short = SA , long = slotted ALOHA}
\DeclareAcronym{SIC}{short = SIC ,long = successive interference cancellation}
\DeclareAcronym{SIR}{short = SIR ,long = signal to interference ratio}
\DeclareAcronym{SNIR}{short = SNIR ,long = signal-to-noise and interference ratio}
\DeclareAcronym{SINR}{short = SINR ,long = signal-to-interference and noise ratio}
\DeclareAcronym{SNR}{short = SNR ,long = signal-to-noise ratio}
\DeclareAcronym{TDM}{short = TDM ,long = time division multiplexing}

\begin{document}

\title{\huge Modern Random Access: an Age of Information Perspective on Irregular Repetition Slotted ALOHA}
\author{Andrea Munari, \IEEEmembership{Senior Member, IEEE}\\
\thanks{A. Munari is with the Inst. of Communications and Navigation, German Aerospace Center (DLR), Oberpfaffenhofen, Germany (email: andrea.munari@dlr.de).}
}
\maketitle
\thispagestyle{empty} \setcounter{page}{0}

\maketitle

\begin{abstract}
Age of information (AoI) is gaining attention as a valuable performance metric for many IoT systems, in which a large number of devices report time-stamped updates to a central gateway. This is the case, for instance, of remote sensing, monitoring, or tracking, with broad applications in the industrial, vehicular, and environmental domain. In these settings, AoI provides insights that are complementary to those offered by throughput or latency, capturing the ability of the system to maintain an up-to-date view of the status of each transmitting device. From this standpoint, while a good understanding of the metric has been reached for point-to-point links, relatively little attention has been devoted to the impact that link layer solutions employed in IoT systems may have on AoI. In particular, no result is available for \emph{modern random access} protocols, which have recently emerged as promising solutions to support massive machine-type communications. To start addressing this gap we provide in this paper the first study of the AoI of a scheme in this family, namely irregular repetition slotted ALOHA (IRSA) \cite{Liva11:IRSA}. By means of a Markovian analysis, we track the AoI evolution at the gateway, prove that the process is ergodic, and derive a compact closed form expression for its stationary distribution. Leaning on this, we compute exact formulations for the average AoI and the age violation probability. The study reveals non-trivial design trade-offs for IRSA and highlights the key role played by the protocol operating frame size. Moreover, a comparison with the performance of a simpler slotted ALOHA strategy highlights a remarkable potential for modern random access schemes in terms of information freshness.
\end{abstract}

\begin{IEEEkeywords}
age of information, grant-free access, irregular repetition slotted ALOHA, IoT, machine-type communications
\end{IEEEkeywords}

\pagestyle{empty}

\newtheorem{prop}{Proposition}
\newtheorem{lemma}{Lemma}
\newtheorem{defin}{Definition}
\newtheorem{remark}{Remark}

\newcommand{\pr}{\ensuremath{\mathbb P}}
\newcommand{\expOp}{\ensuremath{\mathbb E}}

\newcommand{\tru}{\ensuremath{\mathsf S}}
\newcommand{\truIRSA}{\ensuremath{\mathsf S}}
\newcommand{\truSA}{\ensuremath{\mathsf S_{\mathsf{sa}}}}
\newcommand{\psSA}{\ensuremath{\mathsf p_{s,\mathsf{sa}}}}
\newcommand{\load}{\ensuremath{\mathsf G}}

\newcommand{\nodes}{\ensuremath{\mathsf n}}
\newcommand{\slots}{\ensuremath{\mathsf m}}
\newcommand{\pAct}{\ensuremath{\rho}}
\newcommand{\pUpdate}{\ensuremath{\nu}}
\newcommand{\psucc}{\ensuremath{\mathsf p_{\mathsf s}}}
\newcommand{\ploss}{\ensuremath{\mathsf P_{ l}}}
\newcommand{\plosswf}{\ensuremath{\mathsf P_{l,wf}}}
\newcommand{\plossef}{\ensuremath{\mathsf P_{l,ef}}}

\newcommand{\tStamp}{\ensuremath{\sigma^{(i)}}}
\newcommand{\tGen}{\ensuremath{X^{(i)}}}

\newcommand{\age}{\ensuremath{\delta}}
\newcommand{\agei}{\ensuremath{\age^{(i)}}}
\newcommand{\ageSA}{\ensuremath{\age_{\mathsf{sa}}}}
\newcommand{\ageiSA}{\ensuremath{\age^{(i)}_{\mathsf{sa}}}}
\newcommand{\Agei}{\ensuremath{\Delta^{(i)}}}
\newcommand{\avAoI}{\ensuremath{\Delta}}
\newcommand{\avAoIIRSA}{\ensuremath{\avAoI_{\mathsf{irsa}}}}
\newcommand{\avAoISA}{\ensuremath{\avAoI_{\mathsf{sa}}}}
\newcommand{\pv}{\ensuremath{\zeta}}
\newcommand{\pvSA}{\ensuremath{\zeta_{\mathsf{sa}}}}
\newcommand{\pvval}{\ensuremath{\mathsf \theta}}

\newcommand{\actNodesFrame}{\ensuremath{N_a}}
\newcommand{\replicas}{\ensuremath{\ell}}
\newcommand{\tSlot}{\ensuremath{T_{\mathsf s}}}
\newcommand{\tFrame}{\ensuremath{T_{\mathsf f}}}

\newcommand{\slotsRV}{\ensuremath{B}}
\newcommand{\slotsrv}{\ensuremath{b}}

\newcommand{\pGeo}{\ensuremath{\xi}}
\newcommand{\pGeoSA}{\ensuremath{\xi_{\mathsf{sa}}}}

\newcommand{\mcTransProb}{\ensuremath{q}}
\newcommand{\mcTransProbij}{\ensuremath{\mcTransProb_{(i,j)}}}
\newcommand{\mcTransProbGen}[2]{\ensuremath{\mcTransProb_{(#1,#2)}}}
\newcommand{\apRV}{\ensuremath{\Omega}} 
\newcommand{\aprv}{\ensuremath{\omega}}
\newcommand{\apSAk}{\ensuremath{\Psi_{k}}} 
\newcommand{\apSA}{\ensuremath{\Psi}} 
\newcommand{\apsa}{\ensuremath{\psi}} 

\newcommand{\fcomp}{\ensuremath{\beta}}
\newcommand{\scomp}{\ensuremath{\alpha}}

\newcommand{\act}{\ensuremath{\mathcal A}}
\newcommand{\sil}{\ensuremath{\mathcal S}}
\newcommand{\psa}{\ensuremath{\lambda}}
\newcommand{\pas}{\ensuremath{\sigma}}
\newcommand{\pstatA}{\ensuremath{p_\act}}
\newcommand{\pstatS}{\ensuremath{p_\sil}}

\section{Introduction}
\label{sec:intro}

\IEEEPARstart{T}{he} relentless rise of the \ac{IoT} is calling for novel design approaches at all layers of the protocol stack  \cite{Saad20_6G,Munari20_6G}. In particular, the need to provide connectivity to the  vast population of low-power, low-complexity devices encountered in \ac{mMTC} is leading to profound evolutions of medium access protocols \cite{Popovski19_Wiley}. These setups are indeed characterised by transmissions of short packets in a sporadic and unpredictable fashion, which render  traditional grant-based policies highly inefficient due to the excessive cost of overhead.
To effectively support such use cases, \emph{grant-free} access protocols are instead gaining traction. Based on variations of the well-known ALOHA \cite{Abramson:ALOHA}, these approaches represent the de-facto choice for \ac{IoT} in commercial solutions (e.g., LoRa \cite{LoRa}, SigFox \cite{SigFox}) and standards (e.g., NB-IoT, LTE-M \cite{Wang17_CommMag}).

In parallel to this, the need to support massive sets of devices has led to the development of a new family of  protocols, often labelled as \emph{modern random access} \cite{Berioli2016}. These schemes combine novel protocol design ideas with advanced physical layer techniques to go beyond the interference bottleneck of ALOHA. Following different strategies, see, e.g. \cite{Berioli2016,Paolini15:TIT_CSA,Clazzer18:ECRA,Polyanskiy17:RA,Stefanovic13:RatelessAloha,Chamberland20_TIT,Caire19:ISIT,Frolov20_TCOM}, they have been proven to be competitive to grant-based solutions in terms of spectral efficiency, without the need for channel negotiation procedures.  A relevant example is offered by \ac{IRSA} \cite{Liva11:IRSA}, which foresees nodes to independently transmit multiple copies of their packets over a frame of predefined length, and resorts to successive \ac{SIC} to recover information at the receiver side. Thanks to its performance and to the good level of maturity it reached, the scheme became part of the ETSI DVB-RCS2 standard for return-link satellite communications \cite{dvbrcs2}, and is receiving attention as a promising solution in next-generation \ac{mMTC} scenarios \cite{Munari20_6G}.

In parallel to the design of advanced grant-free protocols, the past few years have also witnessed a steadily growing interest towards the definition of novel performance metrics for \ac{IoT}. Many scenarios, in fact, see the presence of a large number of devices which report time-stamped updates to a central gateway. This is the case, for instance, of remote sensing, industrial monitoring or asset tracking, with broad applications in several domains. In these settings, the main goal is often not to optimise spectral efficiency, but rather to provide the gateway with a fresh and up-to-date view of the status of each transmitting device, e.g. to  drive decisions or feedback loops.
These remarks have triggered a florid line of study, aiming at the design of  transmission policies based on the \emph{semantic} of information  \cite{Uysal20_TIT,Ephremides19_AoII,Soleymani20_valueInfo}. The pioneering concept in the field has been that of \ac{AoI}, originally formalised in \cite{Kaul11_SECON,Kaul11_Globecom}. The metric is defined assuming time-stamped messages of equal importance, and captures knowledge freshness at the receiver by measuring the time elapsed since the generation of the latest received update. Despite its simplicity, \ac{AoI} pinpoints a fundamental performance tradeoff, as it depends both on the traffic generation pattern (frequent messages offering fresher information), and on the time spent by an update in the network (higher traffic leading to delivery of older information). As a result, the metric is intrinsically different from classical performance criteria like throughput, as well as from latency, which focuses on  a single packet and only captures its network time.\footnote{Note for instance how latency can often be minimised by lowering the generation rate, reducing waiting times and channel contention. Such an approach can be highly subotpimal for AoI, as producing too sporadic updates results in stale information at the receiver \cite{Yates19_TIT}.} In view of its specificity, AoI has been studied in a number of setups \cite{Modiano19_book}, and a good level of maturity has been reached for point-to-point links, see, e.g. \cite{Yates19_TIT}  and references therein. Interesting results are also emerging in \ac{IoT} setups where multiple devices share the same channel to report to a common destination. Among these, optimal scheduling policies are derived in \cite{Zhou20_TWC} considering packets of different durations to be delivered over noisy channels, whereas joint sampling and orthogonal transmission strategies are studied in \cite{Bedewy19} and in \cite{Zhou19_TCOM} under energy constraints. Distributed and low-complexity scheduling solutions are proposed in \cite{Jiang19_IoT} limiting the negotiation overhead, and in \cite{Modiano19_TNET} meeting throughput requirements, whereas \cite{Ephremides19_Infocom} tackles the behaviour of \ac{AoI} under orthogonal and non-orthogonal multiple access.

On the other hand, research has lately started to focus on models that deal with the additional degree of complexity introduced by having terminals that share the channel via grant-free protocols. Initial and key steps have been taken with the exact characterisation of the average \ac{AoI} of \ac{SA} \cite{Yates17:AoI_SA,Modiano18_AoI}. Both works reveal how maximising throughput also leads to the minimum \ac{AoI} under symmetric channel conditions and for the classical destructive collision model. In the presence of feedback, important improvements were unleashed in \cite{Shirin19_ISIT}, forcing devices to only send updates which would induce a relevant reduction to the AoI. Recent works have also started to look into fully asynchronous random access, deriving \ac{AoI} metrics for pure ALOHA \cite{Yates20_ISIT}.

Despite the lively research on the field, however, the behaviour of \emph{modern random access} from an \ac{AoI} viewpoint is still unexplored, leaving an important gap. In fact, understanding whether solutions which are extremely promising in terms of spectral efficiency are also capable to provide good AoI performance represents a key point to gauge their suitability to a broader set of \ac{mMTC} applications. Motivated by this, the present paper together with its conference version \cite{Munari20_avgAoI}, takes a step to address this relevant question, and provides the first AoI study of an advanced grant-free solution. Specifically, we focus on \ac{IRSA} \cite{Liva11:IRSA}, and consider a setting where a large number of devices send time-stamped status updates to a common destination. In this setup, we provide the following key contributions:
\begin{itemize}
\item resorting to a Markovian analysis, we track the AoI process, prove its ergodicity and derive a compact closed-form expression for its stationary distribution. Leaning on this, we derive the average network \ac{AoI};
\item we then study the behaviour of the age distribution tail, and derive an exact closed-form expression for the probability that the AoI exceeds a threshold. This \emph{age-violation probability} metric provides fundamental insights in many practical IoT settings. This is the case, for instance of dynamic system control in an industrial environment, where decision-making based on stale information may have dire consequences;
\item the Markovian analysis is also applied to a \ac{SA}-based system, leading to simple formulations for average \ac{AoI} \--- in accordance with \cite{Yates17:AoI_SA,Modiano18_AoI} \--- and age-violation probability. A thorough comparison between the performance achievable with the different grant-free access schemes highlights strong gains  offered by modern random access in terms of \ac{AoI};
\item the presented framework reveals non-trivial trade-offs for \ac{IRSA}, for which the closed-form derived expressions offer a simple yet powerful design tool. In contrast to what known for \ac{SA}, we show how tuning the system so as to achieve the maximum throughput is not the only driving force in determining \ac{AoI}. From this angle, the critical role played by the frame duration as well as by IRSA degree distribution is explored and discussed, identifying the best operating conditions for information freshness under different channel loads;
\item the analysis applies to a set of traffic generation profiles for the nodes, providing broadly applicable insights. Furthermore, while the presented results are instantiated for \ac{IRSA}, the analytical framework holds for a larger class of modern random access protocols.
\end{itemize}
The remainder of the manuscript is organised as follows. We start our discussion in Sec.~\ref{sec:sysModel} by introducing the system model considered throughout the discussion, and summarising the key features of the studied grant-free protocols. Starting from this, we derive analytically the behaviour in terms of average and age-violation probability in Sec.~\ref{sec:analysis} for both \ac{IRSA} and \ac{SA}. The performance of the schemes is compared via numerical results in Sec.~\ref{sec:results}, while Sec. ~\ref{sec:burstyTraffic} extends the analysis to a broader class of traffic generation patterns. Finally, conclusions are drawn in Sec.~\ref{sec:conclusions}.

\section{System Model and Preliminaries}
\label{sec:sysModel}

\begin{figure}
  \centering
  \includegraphics[width=.76\columnwidth]{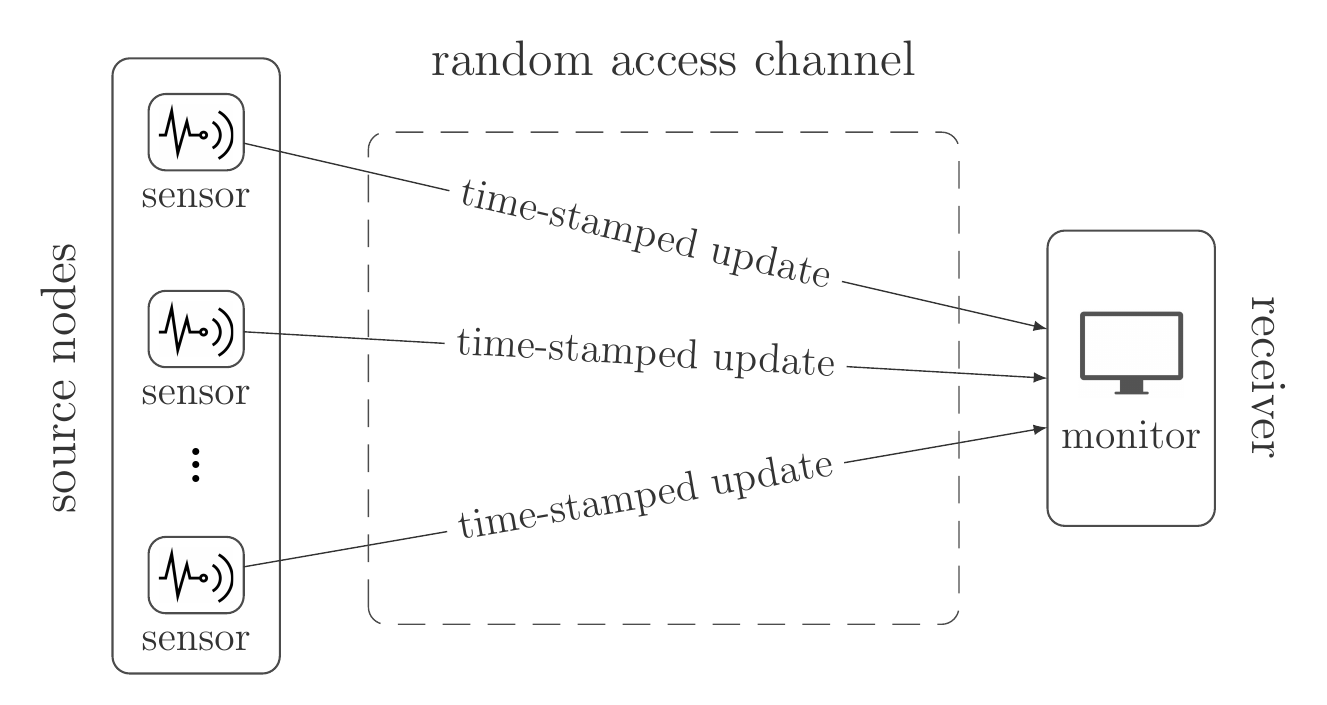}
  \caption{Reference topology: a large population of devices sense a physical quantity and generate time-stamped updates for a common receiver. These are sent over a shared a wireless channel following random access policies.}
  \label{fig:topology}
\end{figure}

We focus on a system with \nodes\ users sharing a wireless channel to attempt delivery of status updates towards a common receiver (sink), as illustrated in Fig.~\ref{fig:topology}. Time is divided in slots of equal duration $T_{s}$, and all terminals are assumed to be synchronised to such pattern. Leaning on this, we measure all time-related quantities in (fraction of) slots, i.e. normalise them to $T_{s}$, and set without loss of generality $T_{s} = 1$.

At each time slot, every user may independently become active with probability \pAct, generating a packet addressed to the sink and marked with a time-stamp that reports the start of the current slot. Such a model is representative of wireless networks in which devices are equipped with sensors to monitor physical quantities, and notify a gateway by sending updates containing a field that specifies the time at which a new measurement was taken. Notable instances of this scenario are given by industrial IoT applications as well as environmental or agricultural MTC settings. A model generalisation to bursty traffic profiles will be presented in Sec.~\ref{sec:burstyTraffic}.

\begin{figure}[!t]
    \centering
    \subfloat[slotted ALOHA (SA)]{
    \includegraphics[width=.8\columnwidth]{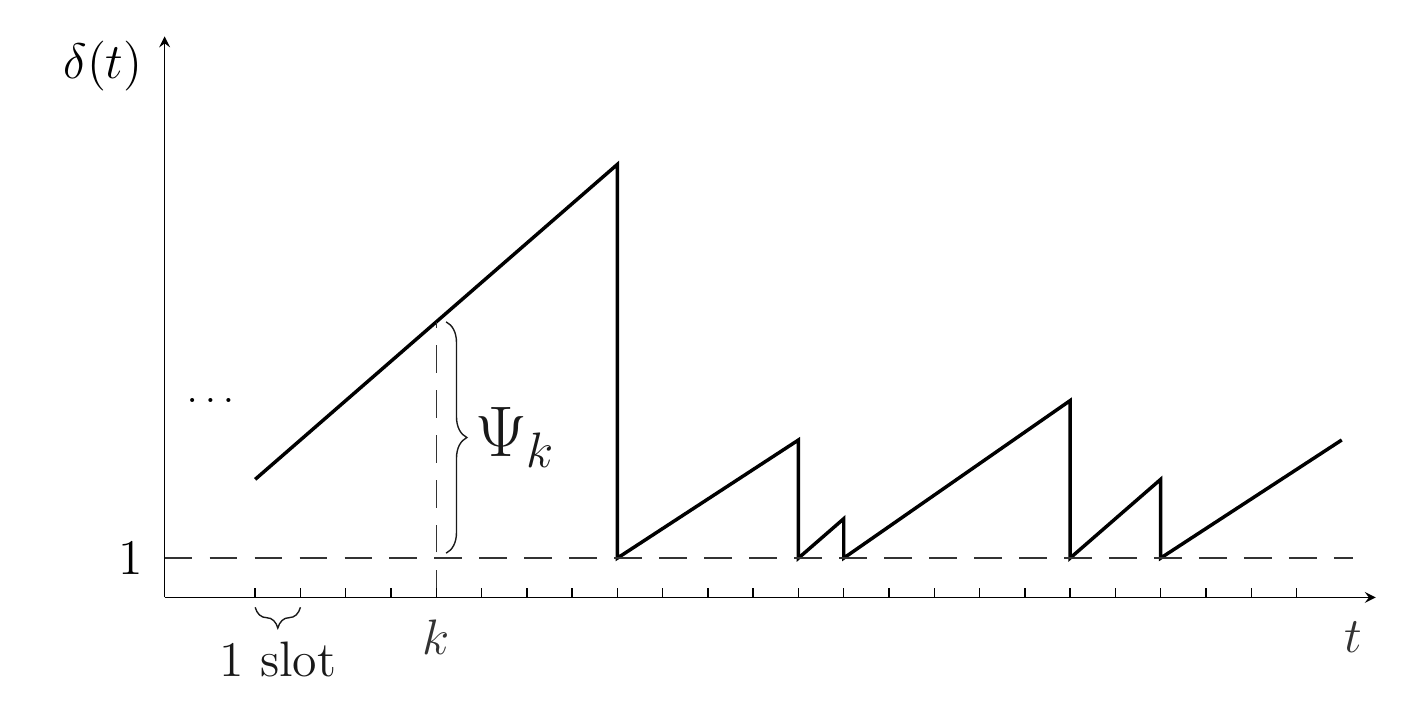}\vspace{-.3em}
    \label{fig:aoiEvolution_sa}
    }\\[-.1em]
    \subfloat[irregular repetition slotted ALOHA (IRSA)]{
    \includegraphics[width=.8\columnwidth]{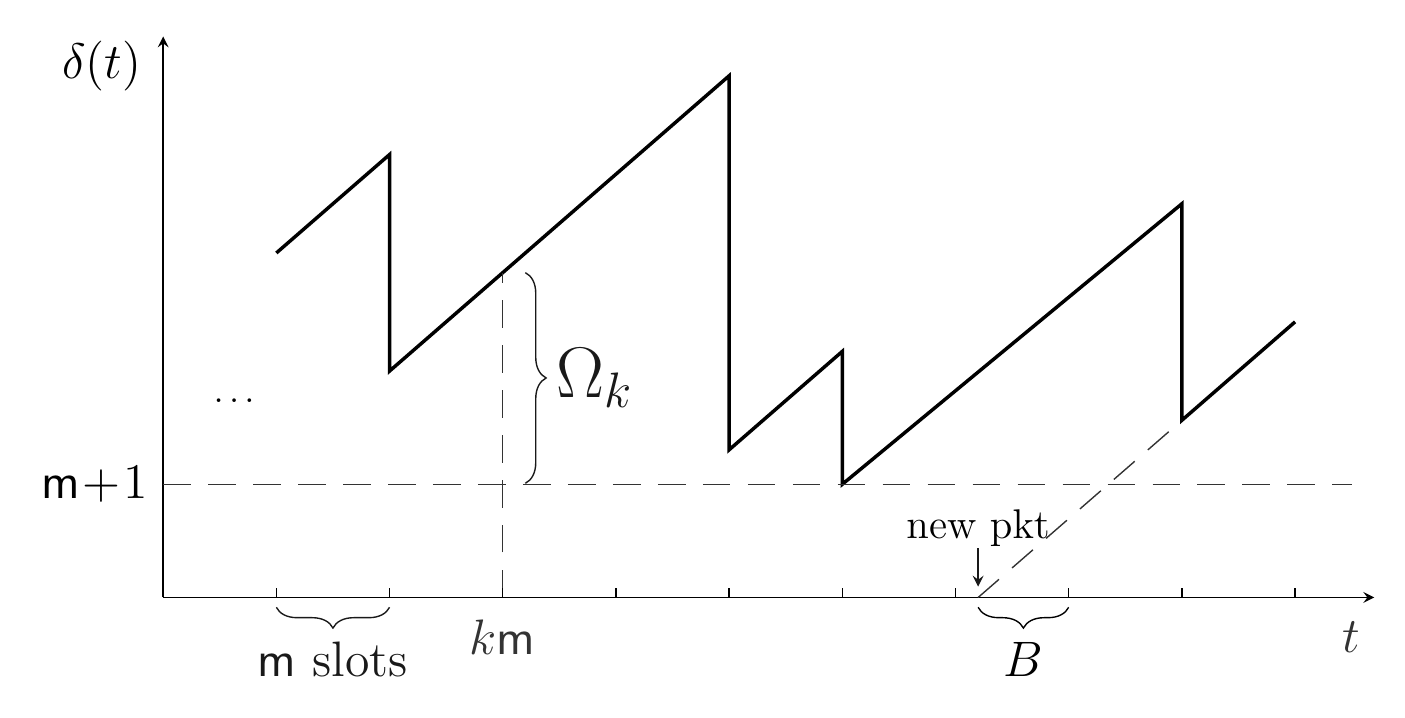}
    \label{fig:aoiEvolution_irsa}
    }
    \caption{Example of realisations for the current \ac{AoI} $\age(t)$ of a source when using \ac{SA} (a), and \ac{IRSA} (b). $\apSAk$ and $\apRV_{k}$ indicate the stochastic processes used to track the current AoI in the two cases (see Sec.~\ref{sec:analysis}).}
    \label{fig:aoiEvolution}
\end{figure}

We shall assume all messages to be equally important, and are interested in having a system that collects and maintains an up-to-date picture of the status of each device. To this aim, we keep track of the \emph{current age of information} for user $i$ at the sink, defined as
\begin{equation}
\agei(t) := t - \tStamp(t)
\label{eq:currentAoI}
\end{equation}
where $\tStamp(t)$ denotes the time-stamp of the last received update from node $i$ as of time $t$.\footnote{Consistently with the definition, also \ac{AoI} related quantities are measured in (fraction) of slots, i.e. normalised to $T_{s}$.} It is relevant to observe that the stochastic process $\agei(t)$ is driven both by the activation pattern of the devices and by the ability to timely deliver updates offered by the employed medium access policy. Specifically, the current \ac{AoI} for a node grows linearly with time, being reset only when a time-stamped packet is decoded. This leads to a typical sawtooth profile exemplified, e.g., in Fig.~\ref{fig:aoiEvolution}. For our study, we focus in particular on the \emph{average network \ac{AoI}}
\begin{equation}
\avAoI:= \frac{1}{\nodes} \sum_{i=1}^{\nodes} \Agei
\label{eq:avgAgeDef}
\end{equation}
where $\Agei$ is the time average of the current \ac{AoI} of the $i$-th node, i.e.
\begin{equation}
\Agei := \lim_{t\rightarrow \infty} \frac{1}{t} \int_{0}^{t} \agei(\tau) d\tau.
\label{eq:avgAgeNode}
\end{equation}
Moreover, we are interested in the long term behaviour of the process, and evaluate the probability for the current \ac{AoI} of a user to exceed a threshold value, as will be formally introduced in Sec.~\ref{sec:analysis} with the notion of  \emph{age violation probability}.

With an eye on \ac{mMTC} applications, we compare the performance of two grant-free access protocols: \ac{SA} \--- taken as a reference benchmark in view of its widespread application \---, and \ac{IRSA}. Following a common and well established modelling assumption for slotted systems, e.g. \cite{Abramson:ALOHA,Paolini15:TIT_CSA}, we regard collisions as destructive, so that the superposition of two or more concurrent transmissions over the same slot prevents decoding at the receiver. Conversely, a slot where a single packet is present always leads to successful decoding. For both schemes, no feedback nor retransmission policies are considered, so that a packet is transmitted at most once.\footnote{We note that the lack of feedback is encountered in many \ac{mMTC} applications, either to reduce downlink bandwidth consumption or to limit devices complexity and energy consumption.} Finally, throughout our discussion we will also refer to the aggregate throughput \tru, defined as the average number of packets per slot decoded at the sink.

\vspace{.3em}
\subsection{Slotted ALOHA}
Following \ac{SA}, a node accesses the channel to send an update as soon as this is generated. Accordingly, at every slot the device successfully delivers a packet to the sink with probability
\begin{equation}
\pGeoSA := \pAct (1-\pAct)^{\nodes-1}.
\label{eq:pGeoSA}
\end{equation}
Here, the first factor $\pAct$ gives the probability that the node becomes active, i.e., has generated a new status update, while the latter grants a successful delivery (i.e. no other user transmits concurrently). In this case, the current \ac{AoI} for the device is reset to $1$, as only the single slot needed for transmission elapses from the generation of the update to its reception (see Fig.~\ref{fig:aoiEvolution_sa}). Since all nodes operate independently, the number of successful deliveries per slot is a binomial r.v. of parameters $(\nodes,\pGeoSA)$, whose expected value gives the aggregate throughput of the scheme: $\truSA = \nodes \pAct (1-\pAct)^{\nodes-1}$.

\begin{figure*}[!t]
    \centering
    \subfloat{
    \includegraphics[width=.365\textwidth]{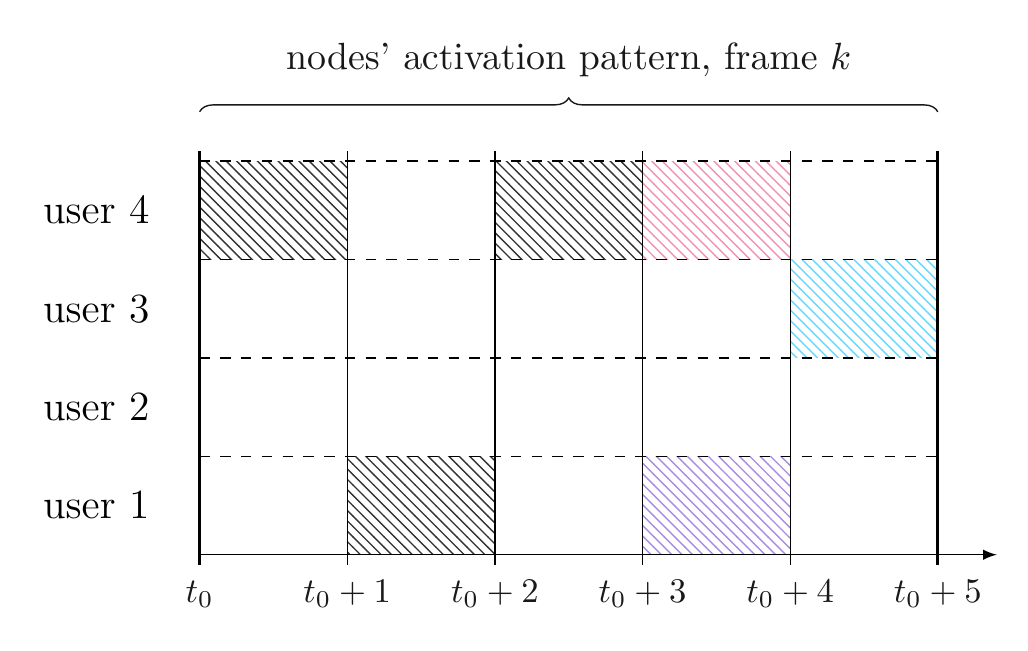}
    \label{fig:irsa_timeline_tx}
    }\hspace{2em}
    \subfloat{
    \includegraphics[width=.365\textwidth]{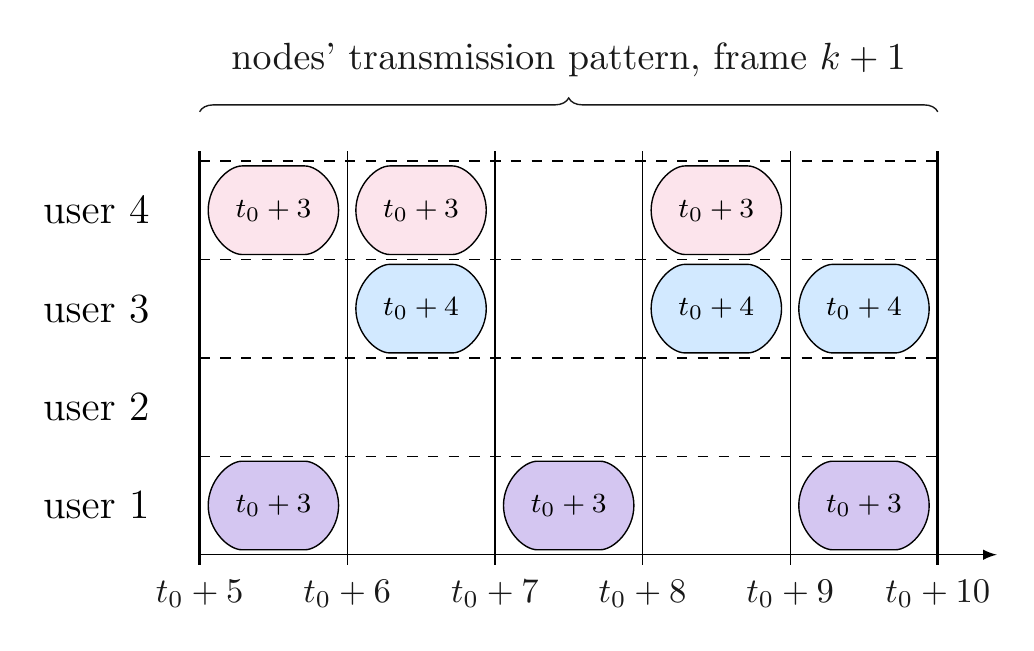}
    \label{fig:irsa_timeline_rx}
    }
    \vspace{-.6em}
    \caption{Example of operation for IRSA with $\nodes=4$, $\slots=5$. On the left handside, the activation pattern of users during a generic frame $k$ is reported, assuming it starts at time $t_{0}$. Filled rectangles indicate slots in which a node became active: note that only the last update (colour fill) will be sent. The right handside shows how nodes access the channel over the subsequent frame $k+1$. Each node that became active transmits three replicas of its update, i.e. $\Lambda(x) = x^{3}$. Every replica contains the time at which the update was created: for instance, user $1$ transmits the (freshest) packet that it generated at time $t_{0}+3$. At the receiver side, processing of frame $k+1$ starts with the decoding of the interference-free second replica of user $1$. Then, the contribution of its twins is removed (SIC), rendering the third copy of the packet of user $3$ decodable. Finally, the update of user $4$ can also be retrieved.}
    \label{fig:irsa_timeline}
\end{figure*}

\vspace{.3em}
\subsection{Irregular repetition slotted ALOHA} \label{sec:sysModel_irsa}
Under \ac{IRSA} operations \cite{Liva11:IRSA}, access to the medium is organised in frames of duration \slots\ slots each. Users are frame-synchronous, and the first transmission opportunity for a new packet is granted only at the start of the next frame. With the considered traffic generation pattern, this implies that a node may become active more than once before being able to access the medium. Aiming to provide the sink with the most up-to-date information, we assume a device in such a situation to transmit the last generated update only, discarding all the others.\footnote{Each node can then be thought of as a one-packet sized buffer, with preemption by fresher updated allowed only in waiting.}

An example of the protocol operations is presented and discussed in Fig.~\ref{fig:irsa_timeline}. As illustrated, a terminal that has become active at least once during an \slots-slot period will initiate a transmission at the start of the subsequent frame. In this case, the node sends $\ell$ identical replicas of its update, uniformly placed at random over the \slots\ available slots. The number of copies is drawn from a probability distribution $\{\Lambda_{\ell}\}$, so that $\ell$ replicas are sent with probability $\Lambda_{\ell}$. Following a common notation inspired by codes on graphs \cite{Paolini15:TIT_CSA,Sandgren17:FACSA,Graell18_Waterfall}, we will compactly denote the distribution using a polynomial formulation as $\Lambda(x) = \sum\nolimits_{\ell} \Lambda_{\ell} \, x^{\ell}$.
Each copy contains a pointer to the positions of its twins, specifying in which slots of the frame they were sent.\footnote{This could be achieved, for instance, by explicitly adding the slot indexes using dedicated header fields. A more efficient alternative which avoids overhead consists in using the payload as seed for a random number generator, used both at the sender and receiver side to place and locate replicas.} At the sink side, the decoding process initiates once the whole frame has been buffered. Specifically, the receiver starts by looking for slots containing a single packet. For any such slot, the corresponding update is decoded. Furthermore, using the pointers contained in the packet, the receiver performs interference cancellation, removing from the incoming waveform the contribution of all replicas of the decoded update. This may in turn lead to more singleton \--- and decodable \--- slots (see Fig.~\ref{fig:irsa_timeline_rx} for an example). The \ac{SIC} procedure is iterated until all users have been retrieved or no more slots with a single packet can be found. For further details, we refer the reader to \cite{Liva11:IRSA}.

For the considered traffic generation profile, the number of users that transmit over a frame follows a binomial r.v. of parameters $(\nodes, 1-(1-\pAct)^{\slots})$. Accordingly, the channel load \load, defined as the average number of devices accessing the channel per slot, is
\begin{equation}
\load = \frac{\nodes \left[1-(1-\pAct)^{\slots}\right]}{\slots}
\end{equation}
and the aggregate throughput can simply be expressed as
\begin{equation}
\truIRSA = \load(1-\ploss)
\label{eq:tru_irsa}
\end{equation}
where \ploss\ denotes the packet loss rate (PLR), i.e. the probability for a packet not to be decoded.

\begin{figure}
    \centering
    \includegraphics[width=.85\columnwidth]{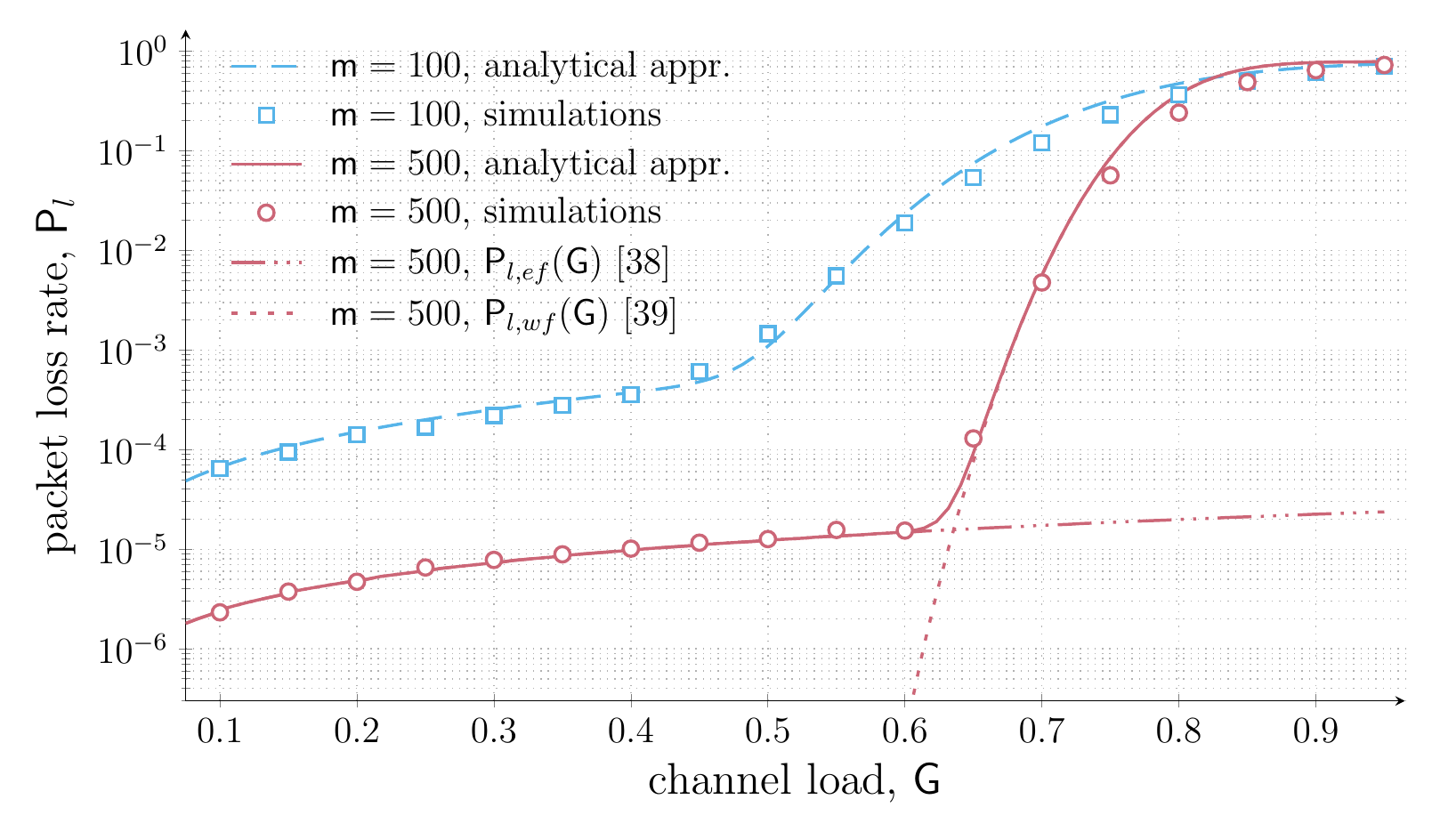}
    \vspace{-.8em}
    \caption{IRSA packet loss rate vs. channel load, assuming $\Lambda(x)=x^{3}$. Solid lines report the analytical approximation in \eqref{eq:plr_approx}, while markers simulations.}
    \label{fig:plr_irsa}
\end{figure}

The computation of \ploss\ would in turn require to track the evolution of the \ac{SIC} process over the finite-length time horizon of a frame, which is a particularly complex task. Due to this, an exact PLR formulation for \ac{IRSA} has proven elusive to date despite numerous research efforts \cite{Sandgren17:FACSA,Graell18_Waterfall}. To characterise the performance of the scheme we then resort to an analytical approximation of \ploss, combining two recent results. The first, introduced in \cite{Sandgren17:FACSA}, captures the behaviour of the protocol in the \emph{error floor} region, i.e. for lightly loaded channels. The second \cite{Graell18_Waterfall}, instead, offers a performance estimate in the \emph{waterfall region}, i.e. for medium to high levels of congestion, adapting the finite length scaling analysis of low-density parity check codes. In both cases, a closed form approximation of the packet loss rate is offered as a function of the channel load, which we denote as $\plossef(\load)$ (error-floor region) and $\plosswf(\load)$ (waterfall region), respectively. The rather involved expressions are given for completeness in App.~\ref{app:appendixIRSA}, and we refer the reader to the original papers for further details. More interestingly, we report in Fig.~\ref{fig:plr_irsa} both $\plossef(\load)$ (dash-dotted line) and $\plosswf(\load)$ (dotted line) for $\Lambda(x)=x^3$ and $\slots=500$ slots, together with the results of Montecarlo simulations (markers). From the plot, it is evident that each expression provides a very good estimate of \ploss\ in one channel load region, and has very little effect on the other one. Leaning on this observation we then approximate the packet loss rate of \ac{IRSA} for a generic channel load as
\begin{equation}
    \ploss(\load) \simeq \plossef(\load) + \plosswf(\load).
    \label{eq:plr_approx}
\end{equation}
Note in fact that, for low values of \load\, \plosswf\ adds a negligible contribution to the sum, whereas it dominates over \plossef\ for larger channel loads.
The good tightness of the approach is confirmed for different frame durations in Fig.~\ref{fig:plr_irsa}, where \eqref{eq:plr_approx} is shown by solid lines. The plot also highlights a well-known behaviour of IRSA, which will be useful to interpret some results in the remainder of the article. We can in fact observe how operating the protocol over longer frames triggers better performance, lowering the error floor and shifting the threshold for entering the waterfall region to higher channel loads \cite{Sandgren17:FACSA,Graell18_Waterfall}.

Let us now focus on the \ac{AoI} evolution experienced by a node when \ac{IRSA} is employed (see, e.g., Fig.~\ref{fig:aoiEvolution_irsa}). Due to the operation mode of the protocol,  $\agei(t)$ grows linearly over a frame, and is reset at the end of it if an update was transmitted and successfully received at the sink, i.e. with probability
\begin{equation}
\pGeo := \big[1 - (1-\pAct)^{\slots}\big] \cdot (1-\ploss) = \frac{\slots \,\truIRSA}{\nodes}.
\label{eq:pGeoIRSA}
\end{equation}
In such a case,  the current \ac{AoI} is refreshed to a value in the set $\{\slots+1, \dots, 2\slots\}$.\footnote{Note that we assume decoding to start only once the whole frame has been stored, and consider the time needed to perform \ac{SIC} over a buffered \slots-slot period to be negligible compared to the duration of a frame, so that it does not affect the \ac{AoI} value.} Indeed, at time of reception, \slots\ slots have elapsed for the transmission process, which add up to the initial offset due to the slots between the generation of the packet and the opportunity for the node to access the channel at the start of the subsequent frame. Throughout our discussion, we describe the latter quantity via the discrete r.v. \slotsRV, for which an example is given in Fig.~\ref{fig:aoiEvolution_irsa}. Recalling the considered traffic generation model, \slotsRV\ takes values in $\{1,\dots,\slots\}$ and has \ac{PMF}
\begin{equation}
p_{\slotsRV}(\slotsrv) = \frac{\pAct(1-\pAct)^{\slotsrv-1}}{1-(1-\pAct)^{\slots}}.
\label{eq:pmfSlots}
\end{equation}
In the equation, the numerator captures the probability for the node to become active \emph{for the last time} at the start of the $(\slots-\slotsrv+1)$-th slot of a frame. For instance, assume that $B = \slots$. This means that the sent packet was generated at the very beginning of a frame, and that the device has not become active again over the remaining $\slots - 1$ slots. The probability for this to happen is exactly $\pAct(1-\pAct)^{\slots-1}$. On the other hand, the denominator in \eqref{eq:pmfSlots} imposes the normalisation condition that at least one packet has to be generated in the frame for the node to access the channel.

\vspace{-1em}
\subsection{Notation} \label{sec:notation}
When analysing \ac{IRSA} operated over frames of \slots\ slots, we shall decompose an integer value $n \in \mathbb N_{0}$ as
\begin{gather}
\begin{split}
n &= \scomp_{n} + \slots \fcomp_{n}\\
\scomp_{n} := \bmod(&n,\slots), \quad \fcomp_{n}  :=\floor*{n/\slots}
\end{split}
\label{eq:alphabeta}
\end{gather}
In other words, we express a generic time slot index $n$ as the sum of $\fcomp_{n}$ frames and of $\scomp_{n}$ additional slots.

Moreover, we indicate by  $\{ \pi_{x}\}$, $x\in \mathbb N$, the stationary distribution of a discrete-time, discrete-valued Markov process $X_{n}$, when it exists. In this case, we refer as the \emph{steady-state} version of the process to a r.v. $X$ with \ac{PMF} $\{ \pi_{x}\}$.

\section{A Markovian AoI analysis for IRSA and SA}
\label{sec:analysis}

We start our study by considering the current \ac{AoI}, introduced in \eqref{eq:currentAoI}, when the system is operated following an \ac{IRSA} access policy. Without risk of confusion we drop the node index $i$, and observe that the stochastic process $\age(t)$  grows linearly within each frame, as its value can only be refreshed at the end of the $\slots$-slot period in case of successful reception of a time-stamped update. Accordingly, we can conveniently write
\begin{equation}
\age(t) = \age\left( \slots\cdot\floor*{t/\slots}\right)+ \bmod(t,\slots)
\label{eq:currAgeDecomp}
\end{equation}
where the first addend captures the age at the beginning of the current frame, while the second accounts for the additional offset from the start of the frame up to the observation time. Without loss of generality, we start tracking the process (i.e., set $t=0$) right after the reception of the first update. Recalling the model introduced in Sec.~\ref{sec:sysModel}, the initial state of our system is thus in the set \mbox{$\{\slots+1, \dots, 2\slots\}$, and $\age(t) \geq \slots + 1$}, $\forall t$. Starting from these remarks, the evolution of the current age is completely specified by tracking the discrete time, discrete valued\footnote{We remark that, while the process $\age(t)$ takes values in $\mathbb R^{+}$, the current \ac{AoI} at the start of a frame can only be equal to an integer number of slots, by virtue of the slotted operation of the protocol described in Sec.~\ref{sec:sysModel_irsa}.} process
\begin{equation}
\apRV_{k} := \age(k\slots) - (\slots + 1), \quad k \in \mathbb N_{0}
\label{eq:apDef}
\end{equation}
which describes the offset of $\age(t)$ at the beginning of the $k$-th frame with respect to its minimum possible value $\slots+1$ (see Fig.~\ref{fig:aoiEvolution_irsa}).

Moreover, since each node operates independently over successive frames, $\apRV_{k}$ is Markovian, and can be described by a chain with state space $\mathbb N_{0}$.
For such process, the one-step transition probabilities
\begin{equation}
\mcTransProbij := \pr\{ \,\apRV_{k+1} = j \, | \, \apRV_{k} = i \,\}
\end{equation}
can readily be computed noting that each frame sees a binary outcome. Specifically:
\begin{itemize}
\item no update is delivered with probability $1-\pGeo$. In such condition the chain simply moves forward from state $i$ to state $j=i+\slots$, tracking the linear increase of the \ac{AoI} over the frame
\item on the other hand, a status update for the node is retrieved with probability \pGeo, given in \eqref{eq:pGeoIRSA}. In this case, the new \ac{AoI} value at the start of frame $k+1$ will be the sum of the $\slots$ slots of frame $k$ (needed to deliver the message with \ac{IRSA}) and the additional slots elapsed between the generation of the update and the start of its transmission process,  described by the r.v. \slotsRV\ introduced in Sec.~\ref{sec:sysModel_irsa}. Accordingly, the Markov chain will transition from state $i$ to state $j=\slotsRV - 1$, $j\in\{0, \dots, \slots-1\}$. For instance, if the successful message originated at the beginning of frame $k-1$, i.e., $\slotsRV = \slots$, the new age value at the start of frame $k+1$ will be $2\slots$, so that the tracked process will be $\slots-1$ slots above its minimum value. Similarly, if the device became active at the last slot of frame $k-1$, we have $\slotsRV = 1$, and the \ac{AoI} will be reset to its minimum value $\slots+1$, so that $\apRV_{k+1}=0$.
\end{itemize}
Leaning on the \ac{PMF} of the r.v \slotsRV\ given in \eqref{eq:pmfSlots}, we can then write the transition probabilities for any state $i\in \mathbb N_{0}$ as
\begin{equation}
\mcTransProbij =
\begin{cases}
\,\pGeo \cdot p_{\slotsRV}(j+1) \quad& j \in \{0, \dots, \slots -1 \}\\
\,1 - \pGeo & j = i + \slots \\
\,0 & \text{otherwise}
\end{cases}
\,.
\label{eq:mcTransProb}
\end{equation}

The formulation in \eqref{eq:mcTransProb} allows to compute the evolution of $\apRV_{k}$ at any point in time, given the initial condition of the system. In general, however, we are more interested in evaluating the long run behaviour of the current age, which we characterise via following result:
\begin{prop} \label{prop:mc}
The stochastic process $\apRV_{k}$ is ergodic, and has steady state distribution\\[-.6em]
\begin{align}
\pi_{\aprv} = \frac{\slots \tru}{\nodes} \left(1-\frac{\slots \tru}{\nodes}\right)^{\!\fcomp_{\aprv}}  \cdot \frac{\pAct (1-\pAct)^{\scomp_{\aprv}}}{1-(1-\pAct)^\slots},\,\,\,\, \aprv \in \mathbb N_{0}
\label{eq:stationaryDist}
\end{align}\\[-.6em]
where $\scomp_{\aprv}$ and $\fcomp_{\aprv}$ are defined as per \eqref{eq:alphabeta}.
\end{prop}
\begin{IEEEproof}
We start by observing that the Markov chain is irreducible, i.e. for any state-pair $(i,j)$, there exists $n>0$ s.t. the $n$-step transition probability between the two states is strictly positive. If $j\in \{ 0\,\dots, \slots-1\}$ the statement holds already at one step from \eqref{eq:mcTransProb}. Otherwise, leaning on \eqref{eq:alphabeta}, let us express $j = \scomp_{j} + \slots \fcomp_{j}$. Then, the transition from $i$ to $j$ can take place in $\fcomp_{j}+1$ steps, moving to state $\scomp_{j} < \slots$ with a successful update, and then experiencing $\fcomp_{j}$ frames without managing to deliver a packet. The event has probability $\mcTransProbGen{i}{\scomp_{j}} \cdot (1-\pGeo)^{\fcomp_{j}} >0$, proving the irreducibility of the chain. Furthermore, observing that each state in $\{0,\dots,\slots-1\}$ has period $1$, the chain is also aperiodic, and admits thus an asymptotic probability distribution $\{\pi_{\aprv}\}$. To compute this, we can write the balance equations. Using the one-step transition probabilities in \eqref{eq:mcTransProb} we obtain:\\[-.3em]
\mathtoolsset{showonlyrefs=false}
\begin{subnumcases}{}
\hspace*{.0em} \pi_{\aprv} = \sum_{j=0}^{\infty} \pi_{j} \cdot \pGeo\, p_{\slotsRV}(\aprv+1) \stackrel{(a)}{=} \pGeo\, p_{\slotsRV}(\aprv+1)  & $\aprv < \slots$ \label{eq:balEqA} \\
\hspace*{.0em}  \pi_{\aprv} = \pi_{\aprv-\slots} \cdot (1-\pGeo)  & $\aprv \geq \slots$ \label{eq:balEqB}\\[-.9em]\nonumber
\end{subnumcases}
\mathtoolsset{showonlyrefs=true}

\noindent where (a) stems from the normalisation condition $\sum\nolimits_{\aprv} \pi_{\aprv} = 1$. For the expressions in \eqref{eq:balEqB}, decomposing the state index as in \eqref{eq:alphabeta} and applying a simple recursion eventually leads to \mbox{$\pi_{\aprv} = \pi_{\scomp_{\aprv}} \cdot (1-\pGeo)^{\fcomp_{\aprv}}$}, where $\pi_{\scomp_{\aprv}}$ is given by the corresponding equation in \eqref{eq:balEqA}. Combining these results, we have, for any $\aprv \in \mathbb N_{0}$
\begin{equation}
\pi_{\aprv} = \pGeo (1-\pGeo)^{\fcomp_{\aprv}} \cdot p_{\slotsRV}(\scomp_{\aprv}+1).
\label{eq:statDistFinal}
\end{equation}
As the Markov chain admits a \emph{proper} stationary distribution ($\pi_{\aprv} > 0$), the process $\apRV_{k}$ is also positive recurrent, and thus ergodic. Plugging the expressions of \pGeo\ and $p_{\slotsRV}(\slotsrv)$ into \eqref{eq:statDistFinal} eventually leads to the statement formulation in \eqref{eq:stationaryDist}.
\end{IEEEproof}

Notably, \eqref{eq:stationaryDist} offers a compact closed-form characterisation of the asymptotic behaviour of the instantaneous \ac{AoI} of a node for any IRSA degree distribution $\{\Lambda_{\ell}\}$.
Leaning on this general result, two quantities of particular interest for system design can be derived: average \ac{AoI} and age-violation probability, which we characterise in the remainder of the section.
\begin{remark}
The derived formulation captures the behaviour of a broader class of traffic generation patterns. Indeed, \eqref{eq:statDistFinal} only requires that i) a node successfully delivers a new update independently over each frame with probability \pGeo; ii) the offset (in slots) between the time stamp of the update and the start of the frame where it is transmitted follows a generic \ac{PMF} $p_B(b)$. For example, assume that a node becomes active over a frame with probability $p_a$, and that the time stamp of the generated update is uniformly distributed. In this case, the packet is delivered with probability $\pGeo=p_a \,(1-\ploss)$, and $p_B(b) = 1/\slots$, $\forall b\in \{1,\dots,\slots\}$. Plugging these into \eqref{eq:statDistFinal} we readily get $\pi_\aprv = \pGeo(1-\pGeo)^{\beta_\aprv}/\slots = (\tru/\nodes)\cdot\left(1-\slots\tru/\nodes\right)^{\beta_\aprv}$. While the formulation in \eqref{eq:stationaryDist} is thus specific to the traffic model described in Sec.~\ref{sec:sysModel}, the presented approach is more general and can easily be adapted to other profiles.
\end{remark}
\begin{figure*}[pb!]
\setcounter{equation}{17}
\normalsize
\hrulefill
\begin{equation}
\pv(\pvval) =
\begin{dcases}
\hspace*{.3em}\frac{\slots \tru}{\nodes} \left(1-\frac{\slots \tru}{\nodes}\right)^{\beta_{\pvval-2\slots}} \cdot \frac{(1-\pAct)^{1+\alpha_{\pvval-2\slots}} - (1-\pAct)^{\slots} }{1-(1-\pAct)^\slots} + \left(1-\frac{\slots \tru}{\nodes}\right)^{1+\beta_{{\pvval-2\slots}}}	&  \pvval > 2 \slots \\
\hspace*{.3em} 1	& \text{otherwise}
\end{dcases}
\label{eq:pv}
\end{equation}
\setcounter{equation}{14}
\end{figure*}

\subsection{IRSA average \ac{AoI}}
A first relevant performance metric is the \emph{average \ac{AoI}}, formally defined in \eqref{eq:avgAgeDef}, for which we provide the following result.
\begin{prop}\label{prop:avgAoI}
The average network \ac{AoI} of \ac{IRSA}, measured in slots and under the traffic model of Sec.~\ref{sec:sysModel}, is given by\\[-.6em]
\begin{equation}
\avAoIIRSA = \frac{\slots}{2} + \frac{\nodes}{\tru} + \left(\frac{1}{\pAct} - \frac{\slots(1-\pAct)^{\slots}}{1-(1-\pAct)^{\slots}} \right).
\label{eq:avgAoIIRSA}
\end{equation}
\end{prop}
\begin{IEEEproof}
Let us focus on a generic node $i$, for which we aim at computing the time average introduced in \eqref{eq:avgAgeNode}, and express the observation interval as $t=k\slots + \bmod(t,\slots)$, i.e. $k$ entire frames plus a portion of the $(k+1)$-th \slots-slot period. Asymptotically, the contribution of the second addend becomes negligible, so that
\begin{align}
\Agei &= \lim_{k\rightarrow \infty} \frac{1}{k\slots} \int_{0}^{k\slots} \agei(\tau) d\tau\\
 &= \lim_{k \rightarrow \infty} \frac{1}{k\slots} \sum_{\ell=0}^{k-1} \int_{\ell \slots}^{(\ell+1)\slots} \!\!\!(\slots+1+\apRV_{\ell} + \tau)\, d\tau.
\end{align}
Here, the second equality follows by: i) expressing the integral as the sum of its components over the $k$ observed frames, and ii) writing the current AoI at time $\tau$ within the $\ell$-th frame as $\agei(\tau)=\apRV_\ell + \slots+1 + \tau$ following \eqref{eq:currAgeDecomp}, \eqref{eq:apDef}. After simple manipulations:
\begin{align}
\hspace*{-1.2em}\Agei \!\!= 1 \!+\! \frac{3\slots}{2} + \lim_{k \rightarrow \infty} \frac{1}{k} \sum_{\ell=0}^{k-1} \apRV_{\ell} \stackrel{(a)}{=} 1 \!+\! \frac{3\slots}{2} \!+\! \sum_{\aprv=0}^{\infty} \aprv \, \pi_{\aprv}
\label{eq:avgNodeAgeStationary}
\end{align}
where (a)  follows with probability $1$ from the ergodicity of the Markov chain proven in Prop.~\ref{prop:mc} \cite{GallagerStochasticProc}. The average \ac{AoI} for the node can then be computed by evaluating the expected value of a r.v. with \ac{PMF} $\{\pi_\omega\}$. Notice that this result is fairly general, as it simply relates the calculation of the area below the curve $\agei(t)$ to the stationary distribution of the process $\apRV_k$, describing the AoI at the start of a frame. Focusing for the moment on the specific traffic profile under consideration, the expression can be further developed using the formulation in \eqref{eq:statDistFinal}. Expressing once more $\aprv \in \mathbb N_{0}$ as $\aprv = \scomp_{\aprv} + \slots \fcomp_{\aprv}$, we have
\begin{align}
\sum_{\aprv=0}^{\infty} \aprv \, \pi_{\aprv} &= \sum_{\fcomp_{\aprv}=0}^{\infty}\sum_{\scomp_{\aprv}=0}^{\slots-1} (\scomp_{\aprv} + \slots \fcomp_{\aprv}) \,\pGeo(1-\pGeo)^{\fcomp_{\aprv}} \, p_{\slotsRV}(\scomp_{\aprv}+1)\\
&= \slots \sum_{\fcomp_{\aprv}=0}^{\infty} \fcomp_{\aprv} \pGeo(1-\pGeo)^{\fcomp_{\aprv}}+ \sum_{\scomp_{\aprv}=0}^{\slots-1} \scomp_{\aprv} p_{\slotsRV}(\scomp_{\aprv}+1).
\end{align}
If we now observe that
\begin{align}
\slots\sum_{\fcomp_{\aprv}=0}^{\infty} \fcomp_{\aprv} \, \pGeo(1-\pGeo)^{\fcomp_{\aprv}} &= \frac{\slots\left(1-\pGeo\right)}{\pGeo} = \frac{\nodes}{\tru} - \slots\\
\sum_{\scomp_{\aprv}=0}^{\slots-1} \scomp_{\aprv} p_{\slotsRV}(\scomp_{\aprv}+1) &= \frac{1}{\pAct} - 1 - \frac{\slots(1-\pAct)^{\slots}}{1-(1-\pAct)^{\slots}}
\end{align}
and plug these values into \eqref{eq:avgNodeAgeStationary}, the right hand-side of the proposition statement follows. Recalling that all nodes in the system operate independently, we have
\begin{equation}
\avAoIIRSA = \frac{1}{\nodes}\sum_{i=1}^{\nodes}\Agei = \Agei
\end{equation}
concluding the proof.
\end{IEEEproof}

\subsection{\ac{IRSA} age-violation probability}
Aiming at an educated system design, the performance of an access scheme in terms of average \ac{AoI} shall be complemented by a characterisation of the distribution tail. Consider, for example, industrial IoT settings in which feedback or actuation are  triggered by the sink based on the latest available sensor measurements. In these scenarios, decisions that rely on stale information shall be avoided, and the worst-case behaviour of the system may be more relevant than its average performance. To explore this aspect, we rely on the \emph{age-violation probability}, originally introduced in \cite{Ephremides14_peakAge}, and recently receiving increased research attention \cite{Durisi19_JSAC,Popovski20_CommLett}. Specifically, in our setup we are interested in evaluating the value assumed by the current AoI $\age(t)$ at the end of a generic frame $k$, just before a possible refresh takes this place. Let us denote this quantity as $\age_k^\prime$. Recalling that the age grows linearly within the frame, this can be expressed as the sum of the AoI value at the start of the frame, i.e. at time $k\slots$, and of the \slots\ slots elapsed in the frame: $\age_k^\prime = \age(k\slots) + \slots$. Accordingly, we introduce the following definition.
\begin{defin}
The \ac{IRSA} \emph{age violation probability}, $\pv(\pvval)$, is the probability for the current AoI value at the end of a frame to exceed the threshold \pvval, in stationary conditions. Formally,
\begin{equation}
\pv(\pvval) := \lim_{k\rightarrow\infty} \pr\{\age_k^\prime > \pvval \}.
\label{eq:def_ageViolation}
\end{equation}
Recalling that $\apRV_k$ is defined as the offset of the current AoI at the start of the $k$-th frame with respect to its minimum possible value $\slots+1$, we can conveniently write
\begin{equation}
\pv(\pvval) = \pr\{ \apRV > \pvval - 2\slots - 1\}
\label{eq:peakAgeIRSA_def}
\end{equation}
where the r.v. $\apRV$ with \ac{PMF} corresponding to the asymptotic distribution of the process $\apRV_{k}$, i.e., $p_{\apRV}(\aprv) = \pi_{\aprv}$, is introduced in accordance to Sec.~\ref{sec:notation}.
\end{defin}
We remark that, thanks to the ergodicity of the Markov process, \pv\ also corresponds to the fraction of the frames in which the receiver experiences an \ac{AoI} above a target value, offering thus and intuitive and useful design tool. Based on this definition, we prove the following result
\begin{prop} \label{prop:pv}
The age-violation probability $\pv(\pvval)$ for IRSA under the traffic model of Sec.~\ref{sec:sysModel} is given by \eqref{eq:pv} reported at the bottom of the page, where the ancillary quantities $\alpha_{\pvval-2\slots}$ and $\beta_{\pvval-2\slots}$ are defined as per \eqref{eq:alphabeta}.
\end{prop}
\begin{IEEEproof}
Recalling that the current \ac{AoI} for a device is never lower than $\slots+1$ slots, and that the value grows linearly over a frame, we readily have $\pv(\pvval) = 1$ for $\pvval < 2\slots + 1$. For larger values, let us introduce for compactness the auxiliary variable $\pvval' := \pvval - 2\slots$. From \eqref{eq:peakAgeIRSA_def}, we then aim to compute $\pr\{\apRV > \theta'\} = \sum_{\aprv>\theta'}\pi_\aprv$. Leaning on \eqref{eq:alphabeta}, let us now write $\pvval' = \scomp_{\pvval'} + \slots \fcomp_{\pvval'}$, i.e. express $\theta'$ as the sum of $\fcomp_{\pvval'}$ frames and $\scomp_{\pvval'}$ additional slots. Similarly, we express $\aprv = \scomp_{\aprv} + \slots\fcomp_\aprv$. Using this decomposition, we observe that $\aprv>\pvval'$ if either: i) \aprv\ contains a larger number of frames than $\pvval'$ (i.e., $\fcomp_\aprv>\fcomp_{\pvval'}$), or ii) $\fcomp_\aprv=\fcomp_{\pvval'}$ and $\scomp_\aprv > \scomp_{\pvval'}$. We can then write
\begin{align}
\pv(\pvval) = \! \sum_{\aprv =\pvval'}^{\infty} \pi_{\aprv} =\!\!\!\sum_{\ell=1+\scomp_{\pvval'}}^{\slots-1} \!\!\!\! \pi_{\ell+\slots\fcomp_{\pvval'}} + \!\!\!\sum_{j=1+\fcomp_{\pvval'}}^{\infty} \sum_{\ell=0}^{\slots-1} \pi_{\ell + \slots j}
\end{align}
where the second equality decomposes the original summation to acocunt for conditions i) and ii).
Plugging in the formulations of \eqref{eq:statDistFinal} and \eqref{eq:pmfSlots},
\begin{align}
\pv(\pvval) = \pGeo (1-\pGeo)^{\beta_{\pvval'}} \cdot \!\!\!\!\sum_{\ell=1+\alpha_{\pvval'}}^{\slots-1} \!\! \frac{\pAct (1-\pAct)^{\ell}}{1- (1-\pAct)^{\slots}}
+ (1-\pGeo)^{\beta_{\pvval'}+1}
\end{align}
observing that $\sum\nolimits_{\ell=1}^{\slots-1} p_{\slotsRV}(\ell) = 1$ and recalling that $\sum\nolimits_{j=1+\fcomp_{\pvval'}}^{\infty} (1-\pGeo)^{j} = (1-\pGeo)^{\pvval'}/\pGeo$. Solving the remaining summation and substituting the value of \pGeo\ derived in \eqref{eq:pGeoIRSA} leads to \eqref{eq:pv}.
\end{IEEEproof}

Before moving to an in-depth discussion of the performance achieved by \ac{IRSA}, two observations are in order. First, it is relevant to stress that the presented framework offers exact closed-form expressions of both average \ac{AoI} and age-violation probability. On the other hand, these quantities are a function of \tru, for which, as discussed, no compact characterisation is known to date. In the remainder of our study we will then evaluate the performance of the protocol resorting to the tight analytical approximation in \eqref{eq:plr_approx}, yet note how the results can be applied to any other (approximated or \--- if available \--- exact) IRSA throughput formulation.

Secondly, the analysis that we propose goes beyond the specific protocol operation of \ac{IRSA}, and can capture the performance of a broader class of schemes. Indeed, equations \eqref{eq:avgAoIIRSA} and \eqref{eq:pv} hold, under the traffic model of Sec.~\ref{sec:sysModel}, for any framed slotted ALOHA approach as long as the proper throughput characterisation is employed. From this standpoint, for instance, the framework also applies to the coded slotted ALOHA family \cite{Paolini15:TIT_CSA}, as well as to T-fold \ac{SA} \cite{Frolov19_tFoldSA}.

\setcounter{equation}{18}
\subsection{Reference Benchmark: slotted ALOHA} \label{sec:analysisSA}
To better gauge the behaviour of \ac{IRSA}, we complement our analysis by deriving the \ac{AoI} statistics for a \ac{SA} policy. In this case, the communication process is not bound to a framed structure, and an update is transmitted as soon as it is generated. For a generic time instant $t$ within the $k$-th slot, i.e. \mbox{$t = k + \bmod(t,k)$}, the current age for a node can be expressed as
\begin{equation}
\ageSA(t) = \ageSA(k) + \bmod(t,k).
\label{eq:ageSA}
\end{equation}
Therefore, the \ac{AoI} evolution can be completely specified by characterising the discrete-time, discrete-valued stochastic process
\begin{equation}
\apSAk := \ageSA(k) - 1, \quad k\in \mathbb N_{0}
\label{eq:apSA}
\end{equation}
which tracks the offset of the \ac{AoI} at the start of slot $k$ with respect to the minimum possible value $1$ (see Fig.~\ref{fig:aoiEvolution_sa}). Also in this case, the process is Markovian, and, recalling \eqref{eq:pGeoSA}, from any state $j$ the chain may either transition to state $0$ with probability $\pGeoSA$, or to state $j+1$ with probability $1-\pGeoSA$. Following a reasoning similar to the one of Prop.~\ref{prop:mc}, the process can readily be shown to be ergodic, and with stationary distribution $\pi_{\apsa} = \pGeoSA(1-\pGeoSA)^{k}$, $k\in \mathbb N_{0}$.

As done for \ac{IRSA}, we can furthermore study the tail of the age distribution by introducing
\begin{defin}
The \ac{SA} \emph{age-violation probability}, $\pvSA(\pvval)$, is the probability that the steady-state version of the current \ac{AoI} for a user exceeds a threshold $\theta$, or, equivalently, the fraction of slots that experience an age above $\pvval$ asymptotically. From \eqref{eq:apSA}, we have
\begin{equation}
\pvSA(\pvval) = \pr\{\apSA > \pvval - 2\}
\end{equation}
where \apSA\ is again introduced for ease of notation as a r.v. having \ac{PMF} $p_{\apSA}(\apsa) = \pi_{\apsa}$.
\end{defin}
Accordingly, we can summarise the behaviour of the protocol with a simple result:
\begin{prop} \label{prop:sa}
For a \ac{SA} system, under the traffic model of Sec.~\ref{sec:sysModel}, the average network \ac{AoI} \--- expressed in slots \--- and the age violation probability are given by\\[-.6em]
\begin{equation}
\avAoISA = \frac{1}{2} + \frac{\nodes}{\truSA}, \quad \,\,
\pvSA(\pvval) =
\begin{cases}
(1-\pGeoSA)^{\pvval-1} & \pvval >1\\
1 & \text{otherwise}
\end{cases}
\label{eq:resultsSA}
\end{equation}
\end{prop}
\vspace{.6em}
\begin{IEEEproof}
The average age can be obtained via simple calculations applying the same approach presented in Prop.~\ref{prop:avgAoI}, whereas the  age violation immediately follows as the complementary CDF of the geometric r.v. \apSA.
\end{IEEEproof}

The presented result for the average \ac{AoI} is in agreement with the formulations originally presented in \cite{Yates17:AoI_SA,Modiano18_AoI}, although obtained in our case from the statistics of the Markov chain. Moreover, to the best of our knowledge, this is the first characterisation of the age violation probability in a \ac{SA} setup.

\section{Results and Discussion}
\label{sec:results}

\subsection{Average age of information}

\begin{figure}
    \centering
    \includegraphics[width=.85\columnwidth]{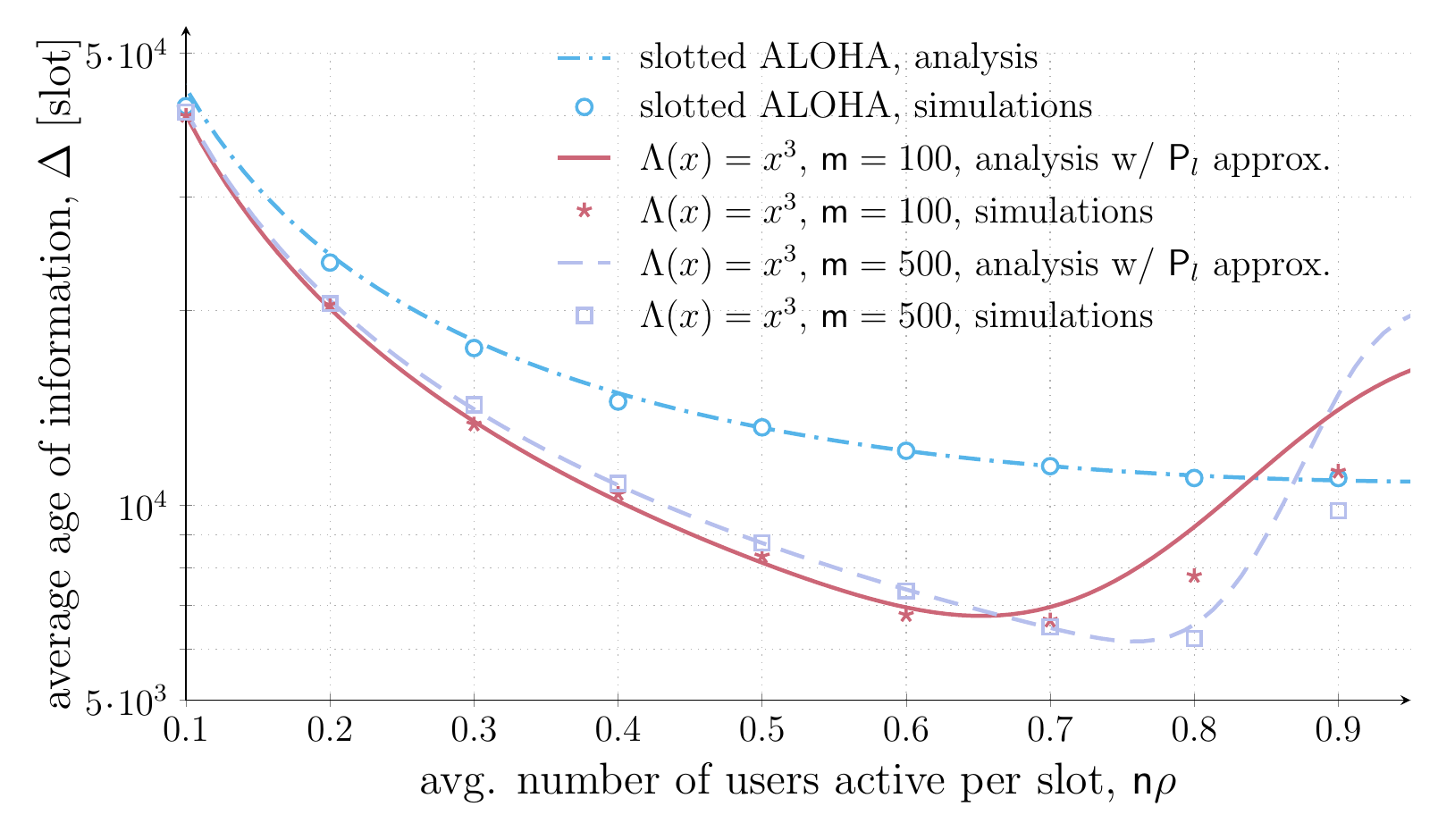}
    \vspace{-.8em}
    \caption{Average network \ac{AoI} (\avAoI) vs. average number of active users per slot ($\nodes\pAct$), for \ac{SA} and \ac{IRSA} ($\slots=100$ and $\slots=500$ slots). All results were obtained for $\nodes=4000$ users and, for IRSA, $\Lambda(x) = x^{3}$.}
    \label{fig:ageVsLoad}
\end{figure}

To draw insights on the behaviour of the considered protocols, we start by studying the average network \ac{AoI}. Unless otherwise stated, we consider a system with $\nodes = 4000$ terminals, and assume a distribution $\Lambda(x) = x^3$ for \ac{IRSA} (i.e., a node accessing the channel sends three copies of its packet over the frame).\footnote{This regular distribution strikes a good balance between time diversity and increased interference due to replicas \cite{Paolini15:TIT_CSA}. Due to this, it has been included in the ETSI DVB-RCS2 standard for the return link of satellite communications systems \cite{dvbrcs2}.} Focusing on this configuration, Fig.~\ref{fig:ageVsLoad} reports the trends of \avAoIIRSA\ and \avAoISA\ as a function of the average number of users per slot that generate an update ($n\pAct$). Markers denote the outcome of Montecarlo simulations.

All access policies exhibit a common behaviour as nodes' activity increases, characterised by a point of minimum for the average \ac{AoI} experienced at moderate-to-high channel load. Indeed, while a lightly loaded medium grants high chance of success for a transmitted update, the behaviour for low values of $\nodes \pAct$ is driven by the scarce activity of terminals, resulting in long inter-update generation times (low \pAct) that penalise \avAoI. Conversely, too frequent reporting from nodes ($\nodes \pAct \sim 1$) lead to channel congestion, hindering decoding of information at the receiver due to collisions.

\begin{table}
\centering
\caption{$\avAoIIRSA/\avAoISA$ for different populations (\nodes) and average number of active users per slot ($\nodes\pAct$). For IRSA, $\slots=500$ and $\Lambda(x) = x^{3}$.}
\label{tab:avgAoIGains}
\begin{tabular}{c|c|c|c}
$\nodes \pAct$ [usr/slot] \Bstrut		& $\nodes = 2000$ 	& $\nodes = 4000$ 	& $\nodes = 6000$ \\
\hline
\hline
$0.2\Tstrut$ 					& 0.8801 			& 0.8494 			& 0.8392	\\
$0.4$ 						&  0.7708			& 0.7206 			& 0.7038	\\
$0.8$ 						&  0.5955			&  0.5879			& 0.6096	\\
\end{tabular}
\end{table}
More interestingly, Fig.~\ref{fig:ageVsLoad} highlights how \ac{IRSA} consistently outperforms \ac{SA} in terms of average \ac{AoI} for most values of \nodes\pAct, and triggers stark improvements exactly in the moderate load conditions under which many practical systems are operated.\footnote{For high $\nodes\pAct$, the trend is reversed. This is due to a well-known behaviour of IRSA \cite{Liva11:IRSA,Paolini15:TIT_CSA}, which offers close-to-linear throughput growth for a broad range of channel loads, prior to suffering a sharp decrease in performance compared to the gentle throughput degradation of SA. In highly congested channels, in fact, the increased interference induced by replica transmission dominates, hindering the gains of \ac{SIC}. The effect reflects on \avAoI\ in view of its inverse proportionality to the throughput. Note however that the system is unlikely to be operated in these conditions, which lead to unacceptably low throughput and reliability.}\footnote{The slight disagreement between simulations and analytical results at high loads stems from the less accurate approximation of the IRSA PLR in this region \cite{Graell18_Waterfall}. We remark that the provided AoI formulation is instead exact.} The trend is buttressed by Tab.~\ref{tab:avgAoIGains}, which summarises the ratio $\avAoIIRSA/\avAoISA$ for different user populations and different channel occupancies. Such a result is non-trivial, as it clarifies how the benefits in terms of throughput efficiency offered by repetitions and \ac{SIC} do outweigh the additional latency cost induced by framed channel access. A first relevant design hint is thus offered, suggesting the use of modern random access solutions as a means to reduce \ac{AoI} in grant-free based \ac{mMTC} applications.

Moreover, Fig.~\ref{fig:ageVsLoad} prompts the natural question on what are the best working conditions (e.g., in terms of $\nodes\pAct$) and how shall systems parameters be tuned in order to optimise \ac{AoI} performance. To tackle this design facet, let us first focus on \ac{SA}. In this case, an inspection of \eqref{eq:resultsSA} readily reveals how \--- for a given terminal population \nodes\ \--- the minimum \ac{AoI} $\avAoISA^*$ is achieved when the aggregate throughput \tru\ is maximised. This happens for $\pAct = 1/\nodes$, leading to
\begin{equation}
    \avAoISA^* = \frac{1}{2} + \nodes \left(1-\frac{1}{\nodes}\right)^{1-\nodes} \simeq
    \frac{1}{2} + \nodes e
    \label{eq:optAoI_SA}
\end{equation}
where the approximation quickly becomes tight for large enough, and practical, values of \nodes. As to \ac{IRSA}, instead, changes in the activation probability \pAct\ affect not only the throughput, but also the additional term accounting for the latency between update generation and transmission (i.e., the third addend in \eqref{eq:avgAoIIRSA}). To further delve into this aspect, let us focus on the typical configurations of \ac{mMTC} applications, characterised by devices that become active in a sporadic fashion (i.e. $\pAct \ll 1$). Under this assumption we can apply the Taylor series expansion $(1-\pAct)^\slots \simeq 1-\slots\pAct + \slots(\slots-1)\pAct^{2}/2$ to \eqref{eq:avgAoIIRSA}, to obtain, after few manipulations, the simpler formulation
\begin{equation}
\avAoIIRSA \simeq \frac{1}{2} + \frac{\nodes}{\truIRSA} + \slots.
\label{eq:avgAoIIRSA_approx}
\end{equation}

\begin{figure}
    \centering
    \includegraphics[width=.85\columnwidth]{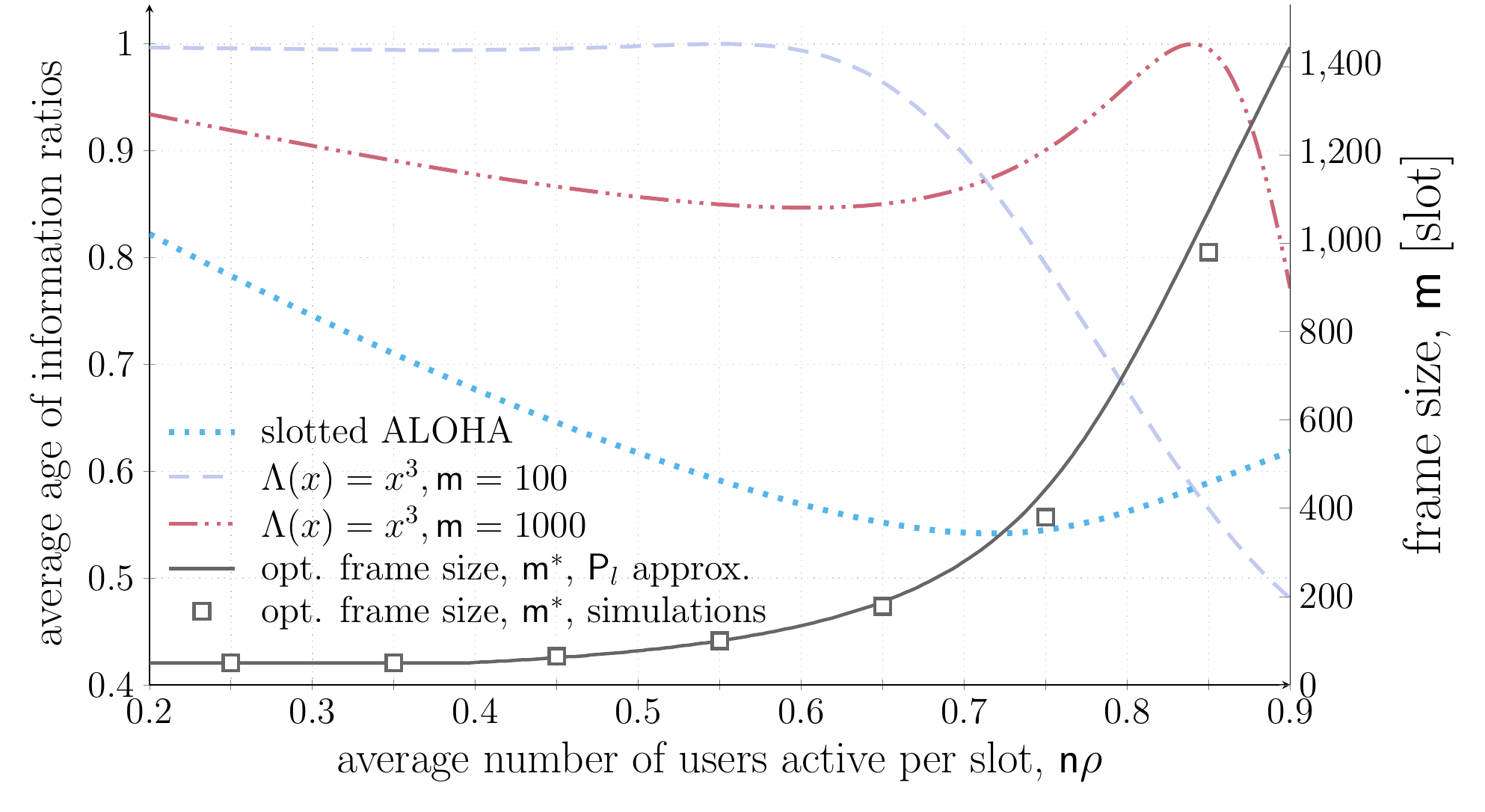}
    \vspace{-.8em}
    \caption{Non-solid lines, referring to the left $y$-axis, report the ratio of the average \ac{AoI} obtained by optimising \ac{IRSA} frame size  to the average \ac{AoI} achieved by other schemes. Specifically, the dashed line indicates the ratio for \ac{IRSA} operated over $\slots=100$ slots, the dash-dotted line for IRSA with $\slots=1000$ slots, andthe  dotted line for \ac{SA}. The solid line, referring to the right $y$-axis, reports the optimal frame size $\slots^{*}$ for \ac{IRSA}. All results obtained for $\nodes=4000$ users and, for \ac{IRSA}, $\Lambda(x)=x^{3}$.}
    \label{fig:frameSizePerf}
\end{figure}
The very tight approximation in \eqref{eq:avgAoIIRSA_approx} allows to draw some interesting remarks. First, it clarifies how, when the frame size is fixed, the average \ac{AoI} is minimised also for \ac{IRSA} by operating the system around its maximum throughput conditions.\footnote{This result may, for instance, be useful in the presence of a feedback channel. In this case, the sink could estimate the experience channel load, and distribute an access probability to the devices so as to improve the \ac{AoI} performance.}  Secondly, the result offers a handy parallel with \ac{SA} \--- for which $\avAoISA = 1/2 + \nodes/\truSA$ (Prop.~\ref{prop:sa}) \---, revealing an additional penalty of $\slots$ slots experienced by IRSA. This term stems from the very operation mode of the protocol, which entails a whole frame duration to transmit and decode a message ($\slots/2$ average \ac{AoI} component already present in \eqref{eq:avgAoIIRSA}), and forces a newly generated update to await the start of the next frame to be sent ($1/2 + \slots/2$ contribution to \avAoIIRSA\ when $\pAct \ll 1$). From this standpoint, \eqref{eq:avgAoIIRSA_approx} pinpoints a fundamental tradeoff for the behaviour of IRSA, and stresses how throughput is not the only force driving AoI. Indeed, while operating over longer frames leads to lower packet loss rates and higher throughput (see Fig.~\ref{fig:plr_irsa}), it also increases the penalty term \slots, possibly worsening the average \ac{AoI}.

\begin{figure}
    \centering
    \includegraphics[width=.85\columnwidth]{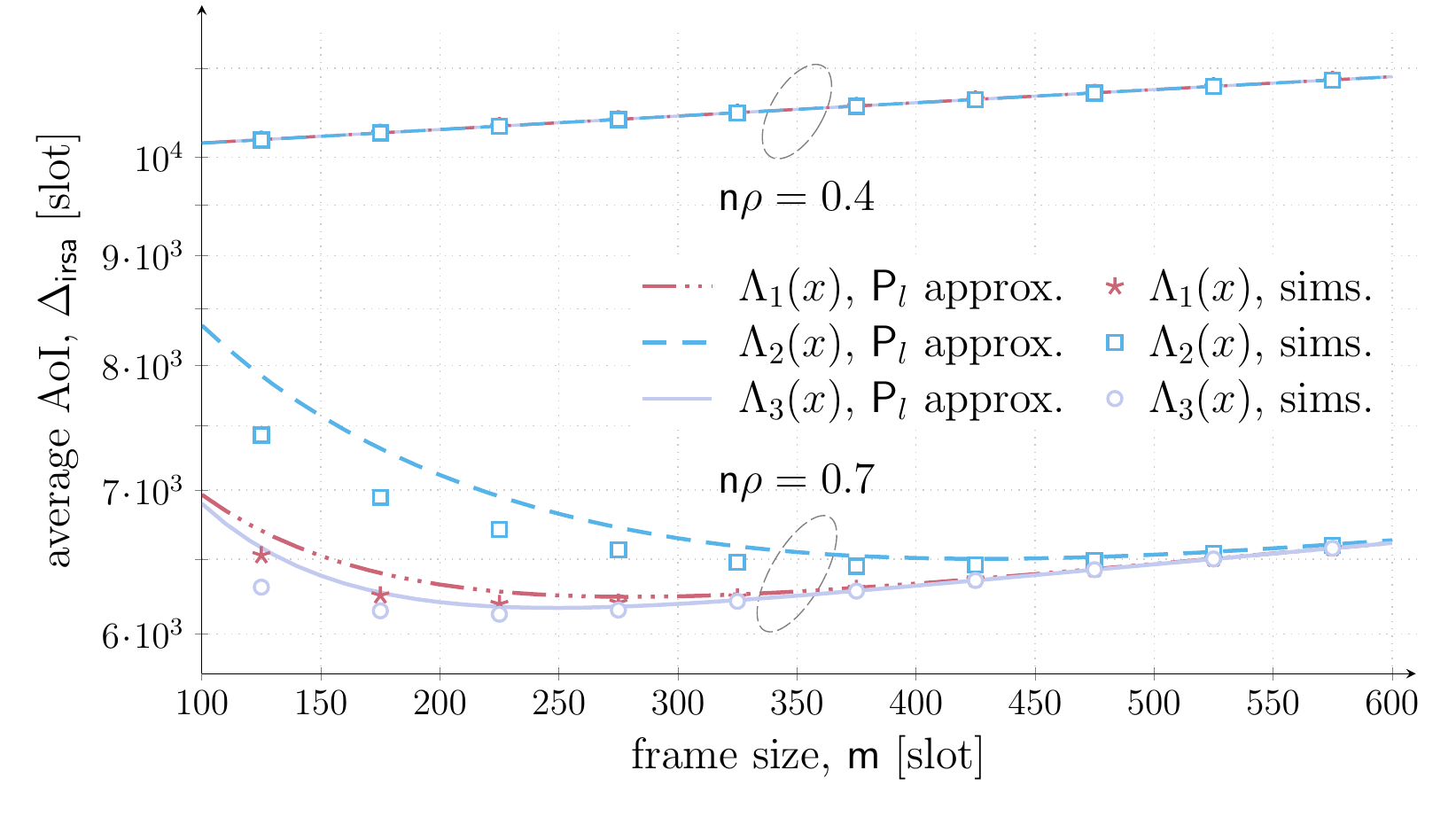}
    \vspace{-.8em}
    \caption{Average network \ac{AoI} (\avAoI) vs. frame size ($\slots$) for different IRSA distributions and different values of $\nodes \pAct$. $\nodes=4000$ users.}
    \label{fig:ageVsFrame_distrib}
\end{figure}
The critical role played by \slots\  is explored in more depth in Fig.~\ref{fig:frameSizePerf}. Specifically, the solid line, whose values are referred to the right $y$-axis, shows the optimal frame length $\slots^{*}$ at which \ac{IRSA} shall be operated for different channel occupancies $\nodes\pAct$ in order to minimise the average \ac{AoI}.
\begin{figure*}[!t]
    \centering
    \includegraphics[width=.8\textwidth]{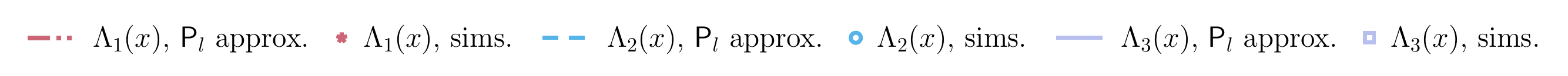}
    \\[-.3em]
    \subfloat[packet loss rate]{
    \includegraphics[width=.28\textwidth]{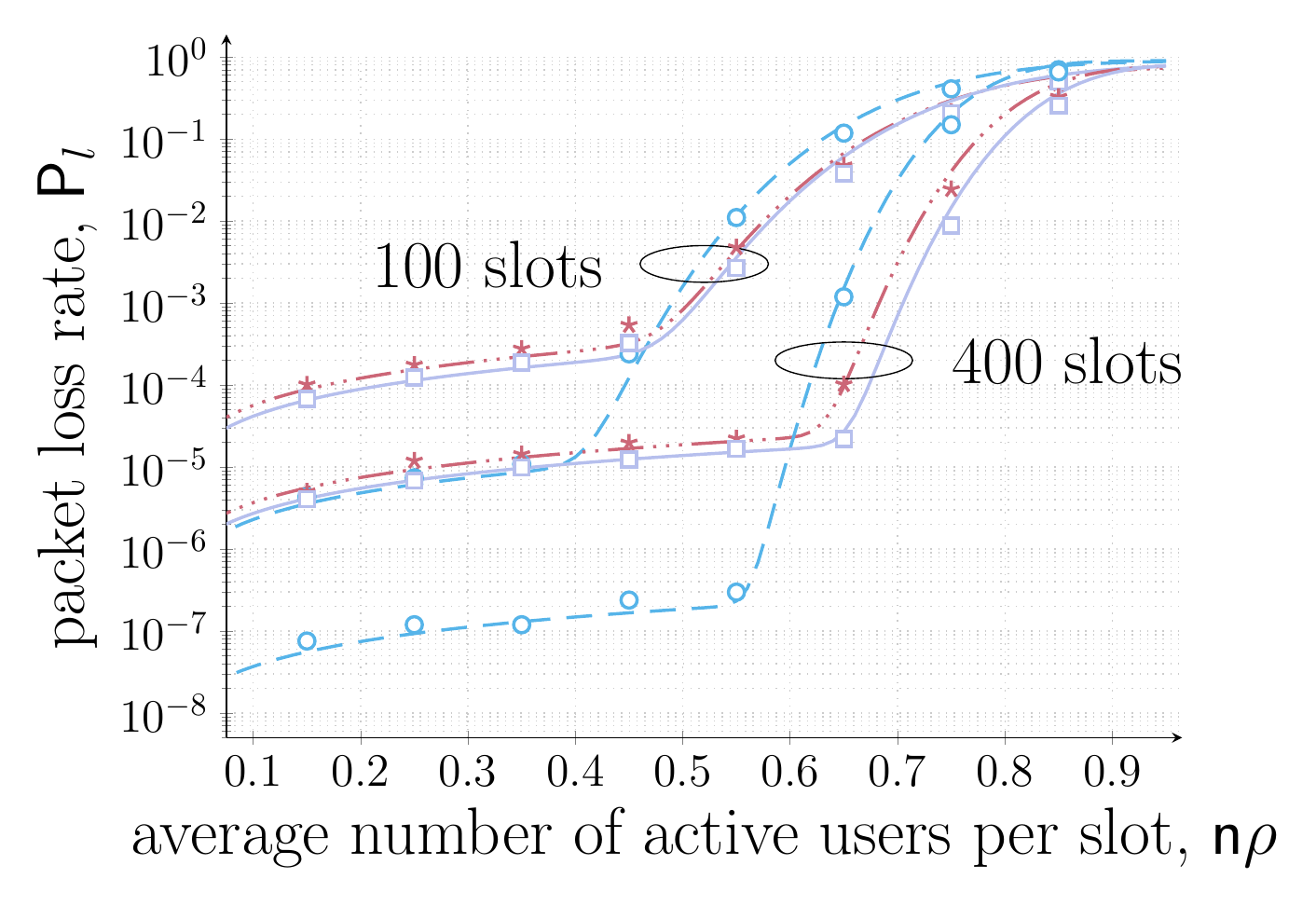}
    \label{fig:plrDiffDistrib}
    }\hspace{1em}
    \subfloat[throughput, $\slots=100$]{
    \includegraphics[width=.28\textwidth]{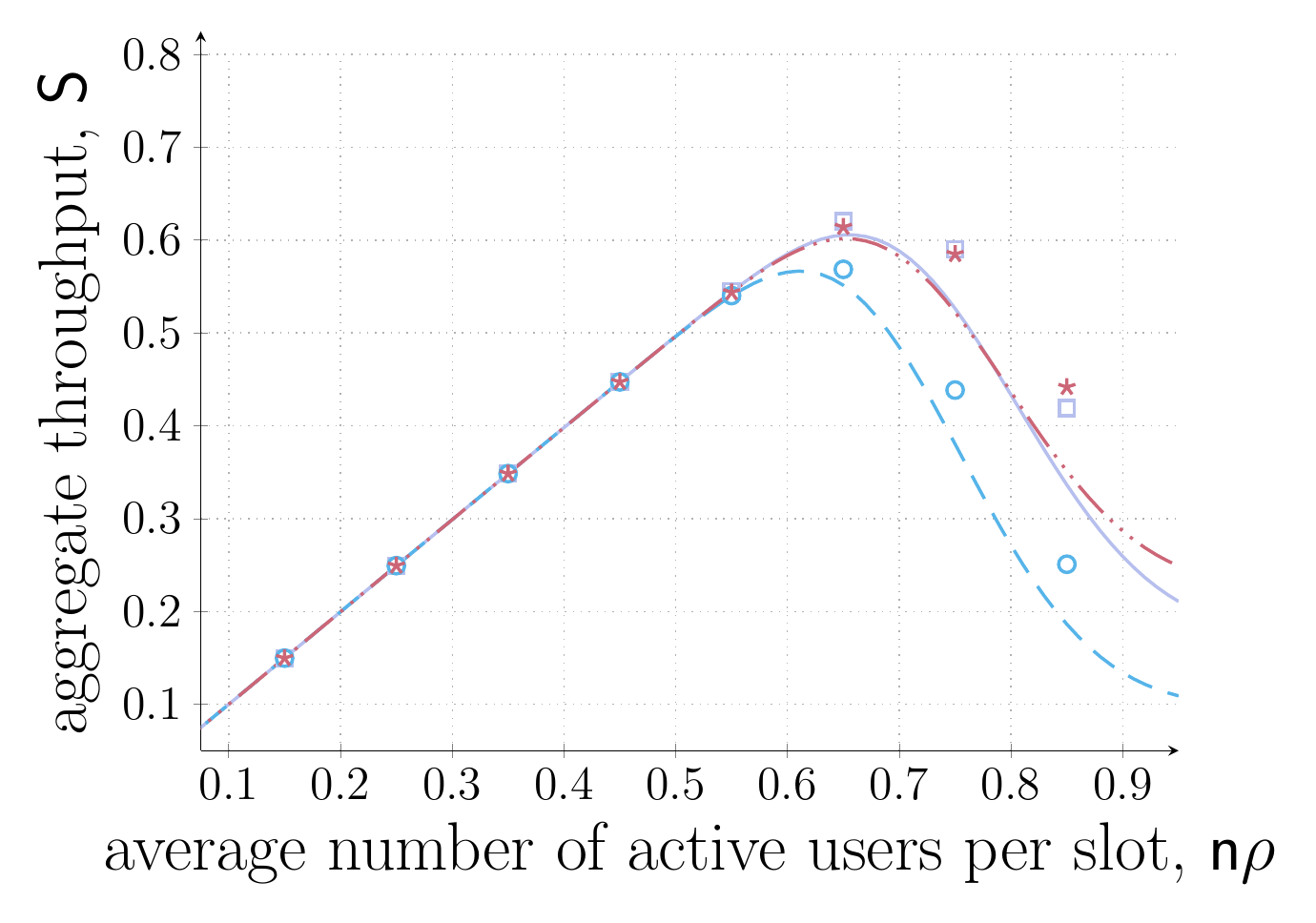}
    \label{fig:truDiffDistrib}
    }\hspace{1em}
    \subfloat[throughput, $\slots=400$]{
    \includegraphics[width=.28\textwidth]{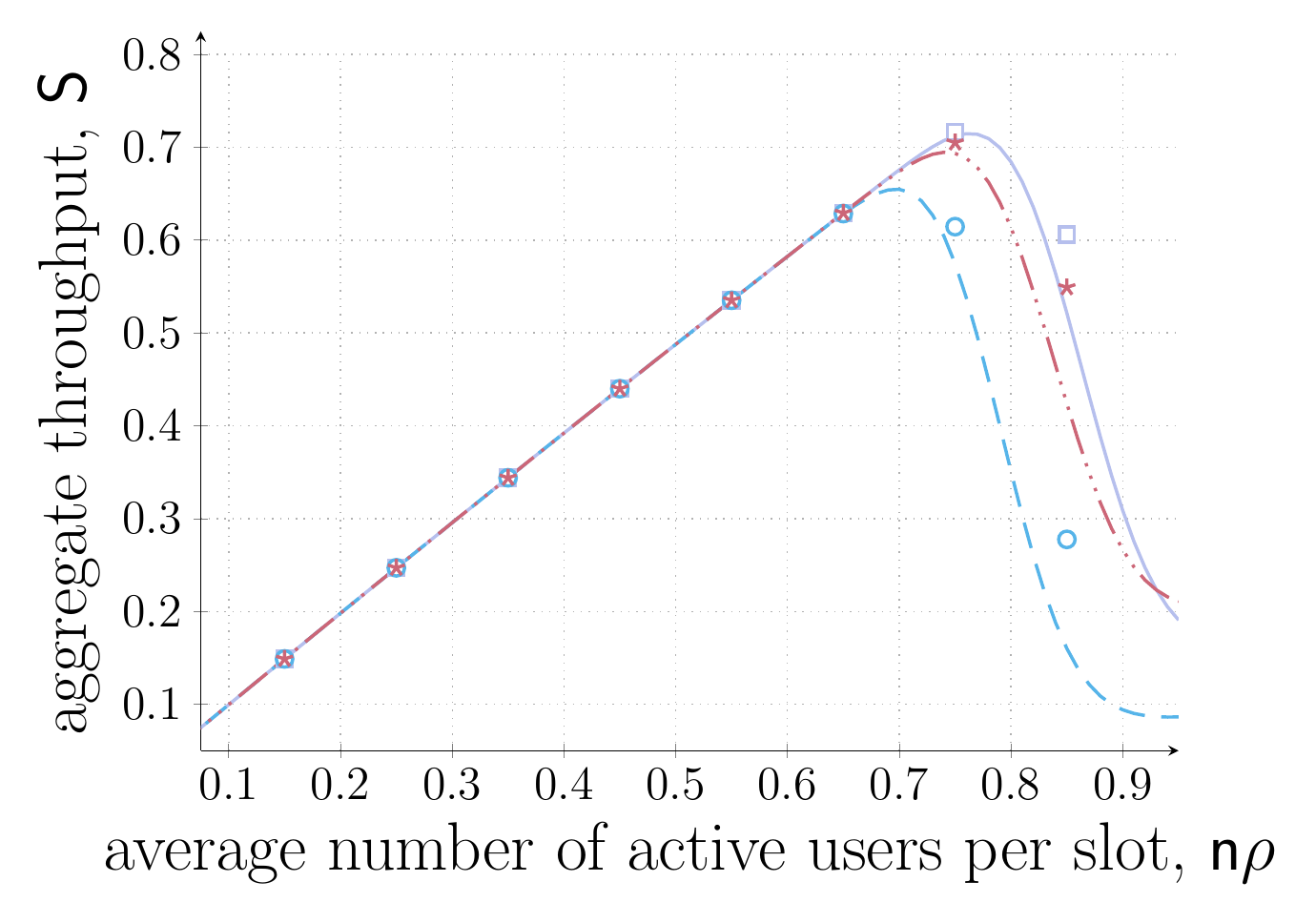}
    \label{fig:truDiffDistrib}
    }
    \caption{Packet loss rate (a) and aggregate throughput ((b) and (c)) vs. average number of active users per slot ($\nodes\pAct$) for different \ac{IRSA} distributions. $\nodes=4000$.}
    \label{fig:plrAndTru}
\end{figure*}
 The trend clearly highlights how small values of \slots\ shall be preferred in lightly loaded channels, while longer frames become beneficial when more devices concurrently access the medium. With low node activity, in fact, IRSA operates in the error-floor region ($\ploss \ll 1$). Here, an increase in frame size does not trigger significant changes in the achievable throughput, rendering \slots\ the driving factor for \avAoIIRSA. Conversely, at higher load, longer frames allow retrieval of a substantially larger number of updates, and the throughput gain more than compensates the additional latency penalty.  In this perspective, it is also interesting to gauge how much benefit a tailored frame duration may bring. The question is tackled once more by Fig.~\ref{fig:frameSizePerf}, where the non-solid lines, referred to the left $y$-axis, specify the ratio of the average \ac{AoI} obtained by optimising \slots\ to the performance achieved otherwise. Specifically, the dashed line indicates such ratio when \ac{IRSA} is operated over a frame of length $\slots=100$ slots, irrespective of the channel load, while the dash-dotted line refers to the case $\slots=1000$. Finally, the dotted line shows the ratio of the optimised \ac{IRSA} average \ac{AoI} to the performance of \ac{SA} (i.e., to $\avAoISA$). The importance of properly selecting \slots\ clearly emerges, becoming especially critical at intermediate channel loads. Indeed, while the loss undergone for employing long frames in lightly loaded channels is rather contained ($\sim 10\%$), operating the system with suboptimal values of \slots\ can lead to stark performance losses when devices become active more frequently. It is also relevant to observe that \--- going beyond the specific outcomes that were illustrated in Fig.~\ref{fig:ageVsLoad} \--- a properly operated version of \ac{IRSA} can outperform \ac{SA} in terms of average \ac{AoI} even for larger values of channel load (e.g., $\nodes\pAct \simeq 0.9$).

All results presented so far were obtained by assuming that each device transmits three copies of its packet over a frame, i.e. relying on $\Lambda_{1}(x)=x^{3}$. To complement our discussion, we further study the impact of using different degree distributions.  Specifically, we focus on the average \ac{AoI} achieved with a simple regular distribution of higher degree, $\Lambda_{2}(x) = x^{4}$, and with the irregular distribution $\Lambda_{3}(x) = 0.86x^{3} + 0.14 x^{8}$, originally derived in \cite{Sandgren17:FACSA} via an optimisation that accounts for performance both in the error-floor and waterfall regions. The behaviour of \avAoIIRSA\ using the different \ac{PMF}s is reported in Fig.~\ref{fig:ageVsFrame_distrib} when changing the frame size \slots\, and for two distinct values of $\nodes\pAct$. The  trends can be properly interpreted recalling the performance induced by the IRSA distribution on packet loss rate and throughput, shown for convenience in Fig.~\ref{fig:plrAndTru} against $\nodes\pAct$ for two relevant frame sizes. Let us first focus on a lightly loaded channel, considering the case $\nodes\pAct=0.4$. In such conditions, all \ac{PMF}s deliver very similar performance. Indeed, \ac{IRSA} is operating in the error-floor region even for short frame sizes ($\slots=100$), so that the beneficial effects on the packet loss rate brought by having users sending more replicas (Fig.~\ref{fig:plrDiffDistrib}) do not translate into substantial throughput \--- and \ac{AoI} \--- improvements (Fig.~\ref{fig:truDiffDistrib}). Instead, the role played by more refined transmission probability distributions kicks in in congested channels, prompting significant differences on the achievable \avAoIIRSA\ (solid curves in Fig.~\ref{fig:ageVsFrame_distrib}, $\nodes\pAct=0.7$). If we compare, for instance, the behaviour of $\Lambda_{2}(x)$ and $\Lambda_{3}(x)$ for $\slots=100$ in Fig.~\ref{fig:plrAndTru}, it is apparent how the former distribution is already operating beyond its throughput peak, whereas the optimised  replica transmission probabilities of the latter enable relevant throughput gains. The effect becomes less pronounced for longer frame sizes, as all the considered distributions tend to perform similarly, leading to the convergence of the average \ac{AoI} curves in Fig.~\ref{fig:ageVsFrame_distrib}. From this standpoint, the plot prompts two further relevant remarks. First, the frame size granting the best performance is tightly related to the employed distribution \--- e.g., for $\nodes\pAct = 0.7$, $\slots \simeq 250$ shall be chosen when using $\Lambda_{3}(x)$, whereas $\slots \simeq 420$ is optimal for $\Lambda_{2}(x)$. Secondly, the notable gains  that emerge were obtained by analysing some simple and well-known distributions, while a dedicated optimisation of the distribution $\Lambda(x)$ focusing on \ac{AoI}-based metrics may further improve performance.

\subsection{Age-violation probability}

Additional system design hints can be derived by looking at the tail of the \ac{AoI} process.
\begin{figure}
    \centering
    \includegraphics[width=.85\columnwidth]{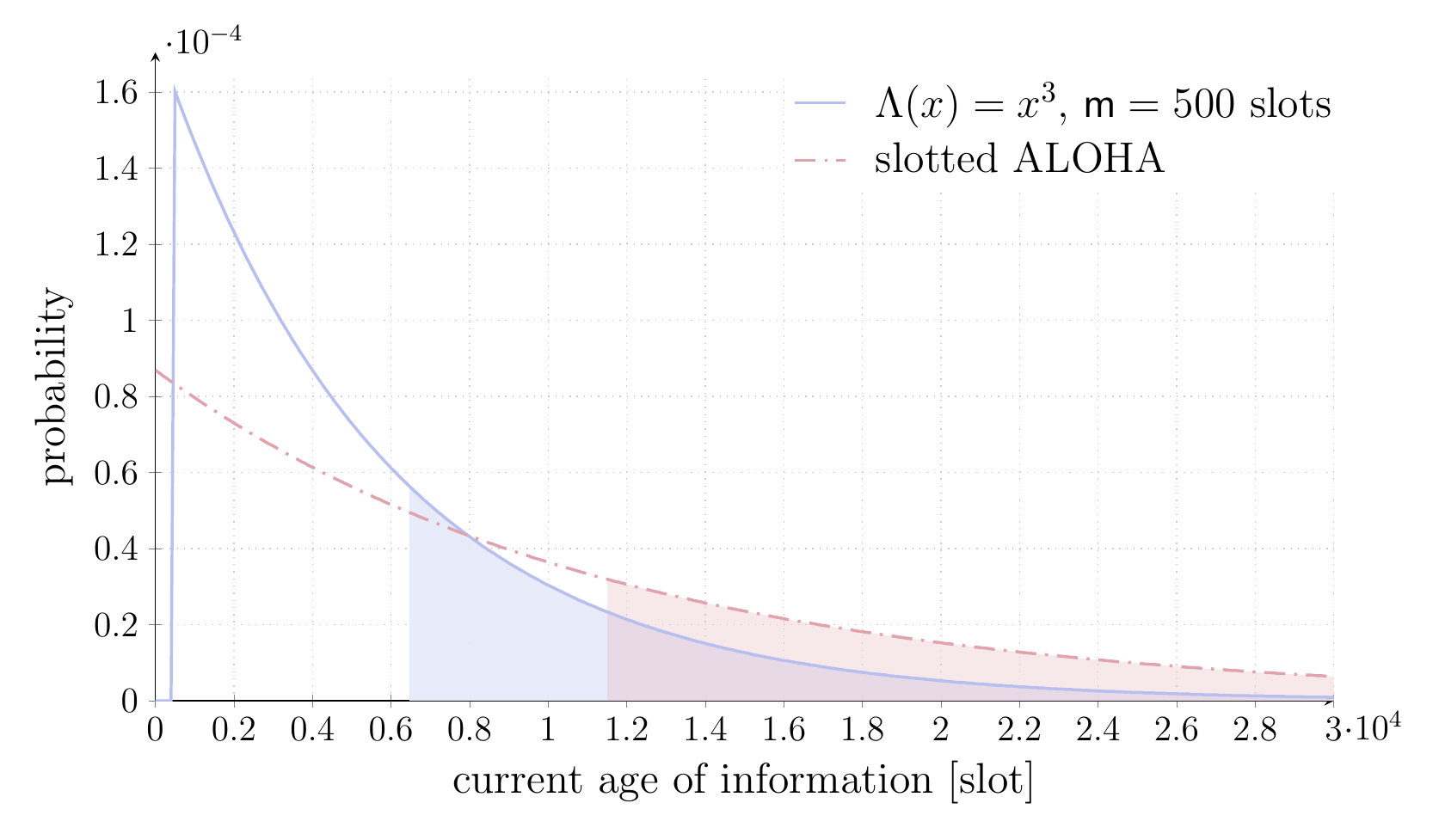}
    \vspace{-.8em}
    \caption{Stationary \ac{PMF} of the current age at the start of a frame for IRSA (solid line) and at the start of a slot for SA (dashed line). For IRSA, $\Lambda(x)=x^{3}$ was considered. In all cases, $\nodes=4000$, $\nodes \pAct = 0.7$.}
    \label{fig:pmf}
\end{figure}
To this aim, Fig.~\ref{fig:pmf} reports for $\nodes\pAct = 0.7$ the stationary distributions of the current age for a node derived in Prop.~\ref{prop:pv} and \ref{prop:sa}. Specifically, the solid curve indicates the age \ac{PMF} at the start of a frame for IRSA (i.e. the distribution of the r.v. $\slots + 1 + \apRV$), while the dashed one the age \ac{PMF} at the start of a slot for \ac{SA} (i.e. the distribution of the r.v. $1 + \apSA$). As expected, \ac{SA} has a non-null probability to experience very low values, whereas the framed operations of \ac{IRSA} bound age to a minimum of $\slots+1$ slots. Conversely, the better packet delivery probability of the modern random access scheme leads to a sharper initial spike of the \ac{PMF}, yielding an overall lower value of average \ac{AoI}.
\begin{figure}
    \centering
    \includegraphics[width=.85\columnwidth]{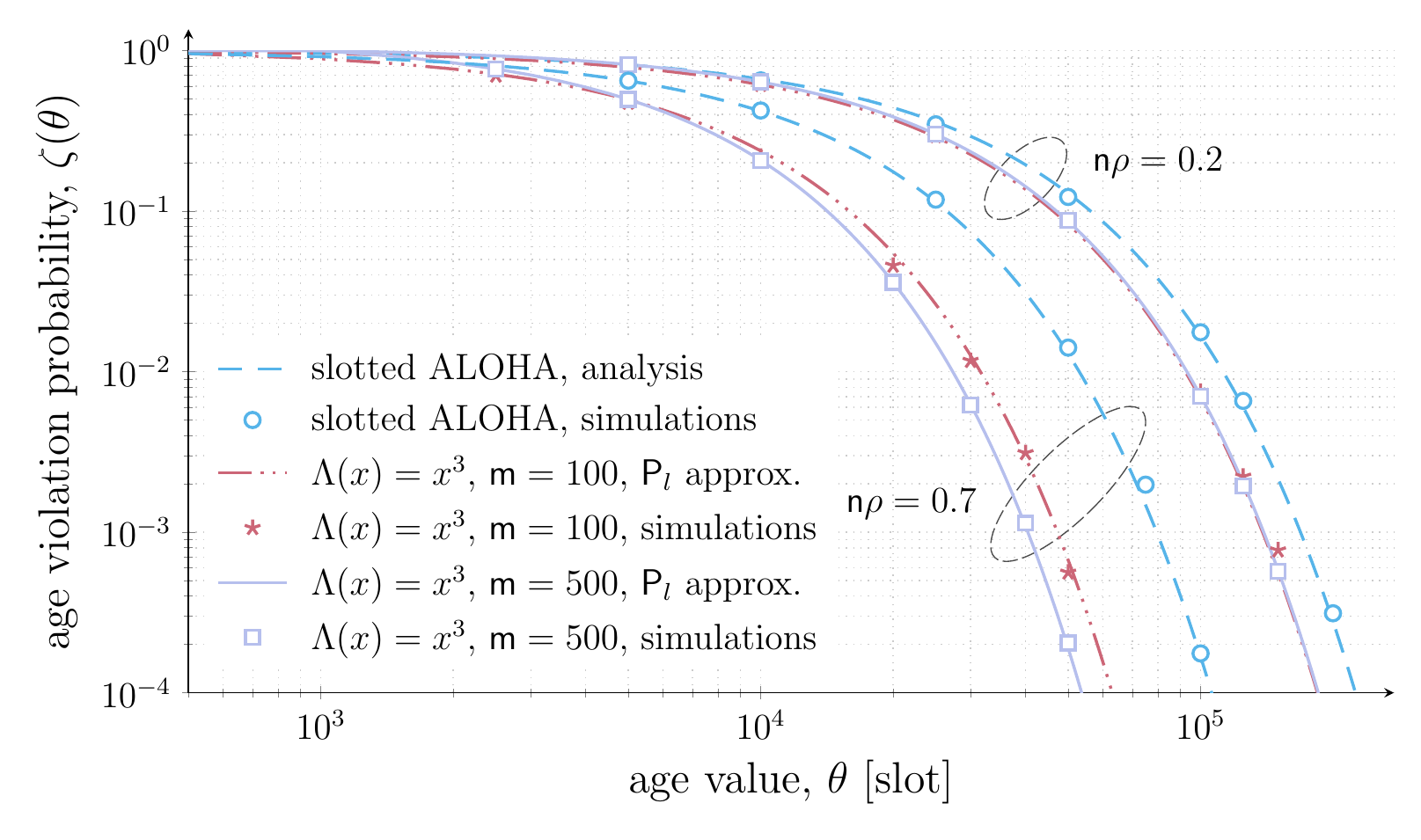}
    \vspace{-.8em}
    \caption{Age-violation probability for slotted ALOHA and IRSA, in lightly-loaded ($\nodes\pAct=0.2$) and more congested  ($\nodes\pAct=0.7$) channel conditions. For IRSA, $\Lambda(x)^{3}$ and, in all cases, $\nodes=4000$.}
    \label{fig:peakAgeViolation}
\end{figure}
More interestingly, the filled areas in the plot highlight values of age that exceed the average for the schemes, revealing in both cases a significant \--- yet differently shaped \--- tail, and stressing the importance of the age violation probability to properly gauge the performance of the protocols. The metric is shown in Fig.~\ref{fig:peakAgeViolation} considering both a lightly loaded ($\nodes\pAct=0.2$) and a more congested ($\nodes\pAct = 0.7$) channel. In all conditions, IRSA outperforms SA, significantly lowering the probability of exceeding a threshold value of tolerable \ac{AoI}. Moreover, the plot pinpoints that longer frames are to be preferred at intermediate to high channel loads from an age-violation probability standpoint as well. On the other hand, only small improvements are attained in lighter traffic conditions by operating the scheme over shorter frames.

\begin{figure}
    \centering
    \includegraphics[width=.85\columnwidth]{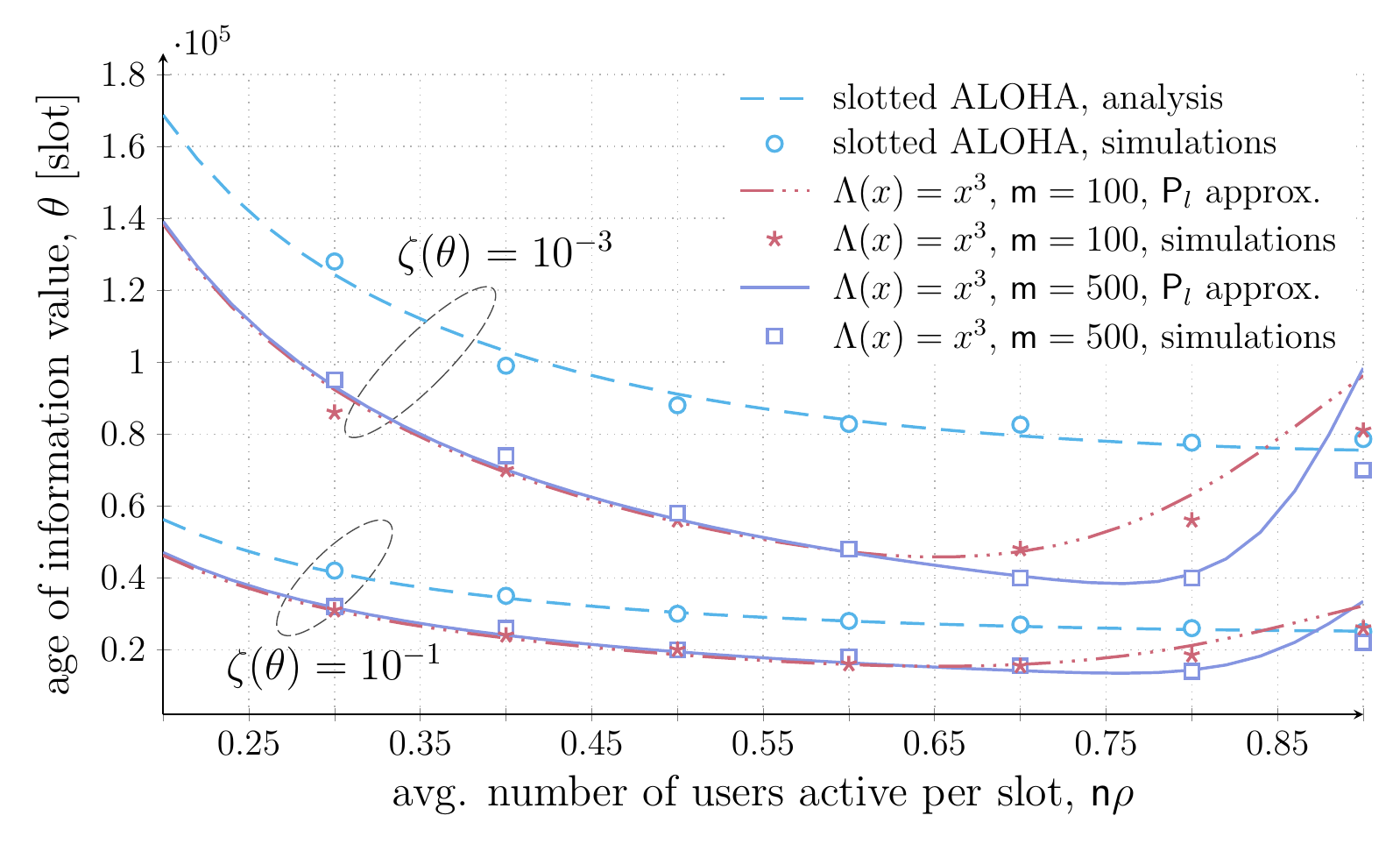}
    \vspace{-.8em}
    \caption{Age of information value for which a target age-violation probability ($10^{-1}$ and $10^{-3}$) is reached for SA and IRSA, reported against $\nodes\pAct$. In all cases, $\nodes=4000$.}
    \label{fig:peakAgeVsLoad}
\end{figure}

These trends are further explored in Fig.~\ref{fig:peakAgeVsLoad}, depicting as a function of $\nodes\pAct$ the \ac{AoI} threshold $\pvval^{*}$ for which a target age violation probability $\pv^{*}$ is reached, i.e. such that $\pv(\pvval^{*}) = \pv^{*}$. In other words, lines report the \ac{AoI} which is exceeded in less than $10\%$ and $0.1\%$ of the frames for IRSA (or slots, for SA). In this perspective, the plot can be seen as a practical tool for system design, allowing an educated choice of the operating parameters based on the desired performance, for which the presented framework offers closed-form expressions. The result confirms how important gains can be achieved by resorting to the modern random access solution, with \pvval\ almost halved compared to SA for channel occupancies or practical relevance (e.g., $\nodes \pAct \simeq 0.7$). Furthermore, in spite of the slightly more complex formulation in \eqref{eq:pv}, the fundamental trade-offs discussed for the average \ac{AoI} are confirmed also when looking at the age distribution tail, showing how good performance can be achieved in high throughput conditions, and stressing the important role played by the selected frame duration.

\section{An Extension of the Framework to Different Traffic Profiles}
\label{sec:burstyTraffic}

As highlighted Sec.~\ref{sec:analysis}, the analysis developed for \ac{IRSA} relies on the assumption that the activation pattern of a node is i.i.d. over successive frames. We now relax this hypothesis, and show how the framework can be applied to a broader set of conditions. Specifically, we will lean on the general expression derived in \eqref{eq:avgNodeAgeStationary}, and derive the average AoI for bursty traffic.

Let us start by defining a more general update generation pattern. We assume that a node can be in one of two states: \emph{active} (\act) or \emph{silent} (\sil). Transitions between the two states can take place at the end of each frame. When silent, the node does not have any packet to deliver. Conversely, in active state the device transmits one status update over the frame. For simplicity, we assume that the update was generated right before the start of the frame where it is transmitted (i.e., $B=1$).\footnote{An extension to a generic distribution $p_B(b)$ is straightforward following the approach of Sec.~\ref{sec:analysis}.} We denote the probability of transition from silent to active as \psa, and the probability of transition from active to silent as \pas. The transmission pattern of a node is then determined by a simple two-state Markov chain. For such a system, it is easy to verify that the stationary probabilities of being in active or silent state are given by $\pstatA = \psa/(\psa+\pas)$ and $\pstatS = \pas/(\psa+\pas)$, respectively.
The considered model is of particular interest for several reasons:
\begin{itemize}
  \item first, it describes well the behaviour of many IoT systems of practical relevance. Indeed, in several applications nodes may monitor a physical quantity of interest, and trigger an update transmission only if an event is detected (e.g., exceeding a threshold). In these cases, periods of silence may be followed by a burst of transmissions once the node becomes active, as captured by the two-state Markov chain;
  \item the possibility to transmit with persistency ($1-\pas$) over successive frames may also be representative of a simple link layer strategy to attempt reducing AoI. Assume for instance that no feedback from the sink is provided, as typical in many IoT scenarios. In this case, a device that wants to report to the destination may decide to generate and transmit new updates over a sequence of successive frames, in the hope that at least one of the packets is decoded at the sink. This repetition-based approach, followed for example by the Sigfox solution for IoT systems \cite{SigFox}, triggers however a trade-off. In fact, the increased number of transmission may lead to higher channel congestion and thus reduce the success probability, with detrimental effect on \ac{AoI}. As will be discussed later, the considered model allows to evaluate the effectiveness of the strategy;
  \item finally, the two-state model represents an extension of the system model described in Sec~\ref{sec:sysModel}. Indeed, setting $\psa = p_a$ and $\pas = 1-p_a$ leads back to a system in which the transmission pattern of a node is i.i.d. across frames with probability $p_a$. The results discussed in the first part of the paper can thus be seen as obtained for a specific configuration of the traffic profile being described.
\end{itemize}
To study the behaviour of the system, we start again by looking at the evolution of the AoI measured at the beginning of an IRSA frame. More precisely, we are interested in deriving the stationary distribution of the stochastic process $\apRV_k$, which, we recall, is the offset of the current AoI at the start of frame $k$ with respect to its minimum possible value $\slots+1$ (see also Fig.~\ref{fig:aoiEvolution_irsa}).
To achieve this goal, we now need to \emph{jointly} track the evolution of $\apRV_k$ and of the activity of the source $s_k \in \{\act, \sil\}$, since the probability of sending an update over successive frames is no longer independent. We therefore look at the Markov chain with state $(\apRV_k, s_k)$, whose one-step transition probabilities can easily be derived:
\begin{itemize}
  \item from state $(0,\sil)$ the chain can transition to state $(\slots,\act)$ with probability $\psa$, or to state $(\slots,\sil)$ with probability $1-\psa$. In fact, if the node is silent, no update will be transmitted, and the AoI will linearly grow over the frame, increasesing by $\slots$ slots;
  \item similarly, from state $(\ell\slots,\sil)$, $\ell > 0$, the chain can transition either to state $((\ell+1)\slots,\act)$, with probability $\psa$, or to state $((\ell+1)\slots,\sil)$, with probability $1-\psa$;
  \item finally, from state $(\ell,\act)$, $\ell \geq 0$, the chain will move to state $(0,\act)$ with probability $(1-\ploss)(1-\pas)$. This happens when the sent update is successfully received, factor $(1-\ploss)$, and the node remains active, factor $(1-\pas)$. Along the same line, the chain transitions to state $(0,\sil)$ with probability $(1-\ploss)\pas$. If, instead, the transmitted update is lost, the chain moves to state $((\ell+1)\slots,\act)$ \--- with probability $\ploss (1-\pas)$ \--- or to state $((\ell+1)\slots,\sil)$, with probability $\ploss\,\pas$.
\end{itemize}
The Markov process is once more ergodic, and its stationary behaviour can be determined by writing the balance equations. Denoting as $\pi_\ell^{\act}$ and $\pi_\ell^{\sil}$ the stationary probabilities of being in state $(\ell\slots, \act)$ and $(\ell\slots,\sil)$, $\ell \geq 0$, respectively, and recalling the described transition probabilities, we have
\begin{align}
  \begin{dcases}
  \pi_{0}^\sil = \sum_{\ell=0}^\infty \pi_{\ell}^\act \,(1-\ploss)\,\pas \stackrel{(a)}{=}  p_\act \cdot (1-\ploss) \,\pas &\\
  \pi_{0}^\act = \sum_{\ell=0}^\infty \pi_{\ell}^\act \,(1-\ploss)(1-\pas) \stackrel{(b)}{=}  p_\act \cdot (1-\ploss) (1-\pas) &\\
  \pi_{\ell}^\sil = \pi_{\ell-1}^\sil \cdot (1-\psa) + \pi_{\ell-1}^\act \cdot \ploss\,\pas & \hspace{-1.2em}\ell > 0 \\[.2em]
  \pi_{\ell}^\act = \pi_{\ell-1}^\sil \cdot \psa \,+\, \pi_{\ell-1}^\act \cdot \ploss (1-\pas) & \hspace{-1.2em} \ell > 0 \\[-1em]
  \label{eq:balanceEquatinosBursty}
  \end{dcases}
\end{align}
Here, (a) and (b) follow from the fact that  $\sum_\ell \pi_\ell^\act$ gives the stationary probability of being in active state, $p_\act$, computed at the beginning of this section. In contrast to the derivation of Sec.~\ref{sec:analysis}, a closed form solution for the system is unfortunately elusive. However, \eqref{eq:balanceEquatinosBursty} readily leads to the matrix relation
\begin{align}
  \begin{pmatrix}
    \pi_\ell^\sil\\
    \pi_\ell^\act
  \end{pmatrix}
  &= \begin{pmatrix}
    1-\psa  & \ploss\,\pas\\
    \psa    & \ploss (1-\pas)
  \end{pmatrix}^{\!\!\mbox{\small$\ell$}}
  \begin{pmatrix}
    \pi_0^\sil\\
    \pi_0^\act
  \end{pmatrix}\\
  &=
  \begin{pmatrix}
    1-\psa  & \ploss\,\pas\\
    \psa    & \ploss (1\!-\!\pas)
  \end{pmatrix}^{\!\!\mbox{\small$\ell$}}
  \begin{pmatrix}
    \pas\\
    1\!-\!\pas
  \end{pmatrix} \frac{\psa(1-\ploss)}{\psa + \pas}\\[-1.6em]
  \label{eq:matrixSolutionBursty}
\end{align}
which allows an easy numerical computation of $\pi_\ell^\sil$ and $\pi_\ell^\act$ for any $\ell \geq 0$. Finally, the sought stationary distribution of $\apRV_k$ can simply be obtained as $\pi_\ell = \pi_\ell^\sil + \pi_\ell^\act$.
\begin{figure}
  \centering
  \includegraphics[width=.85\columnwidth]{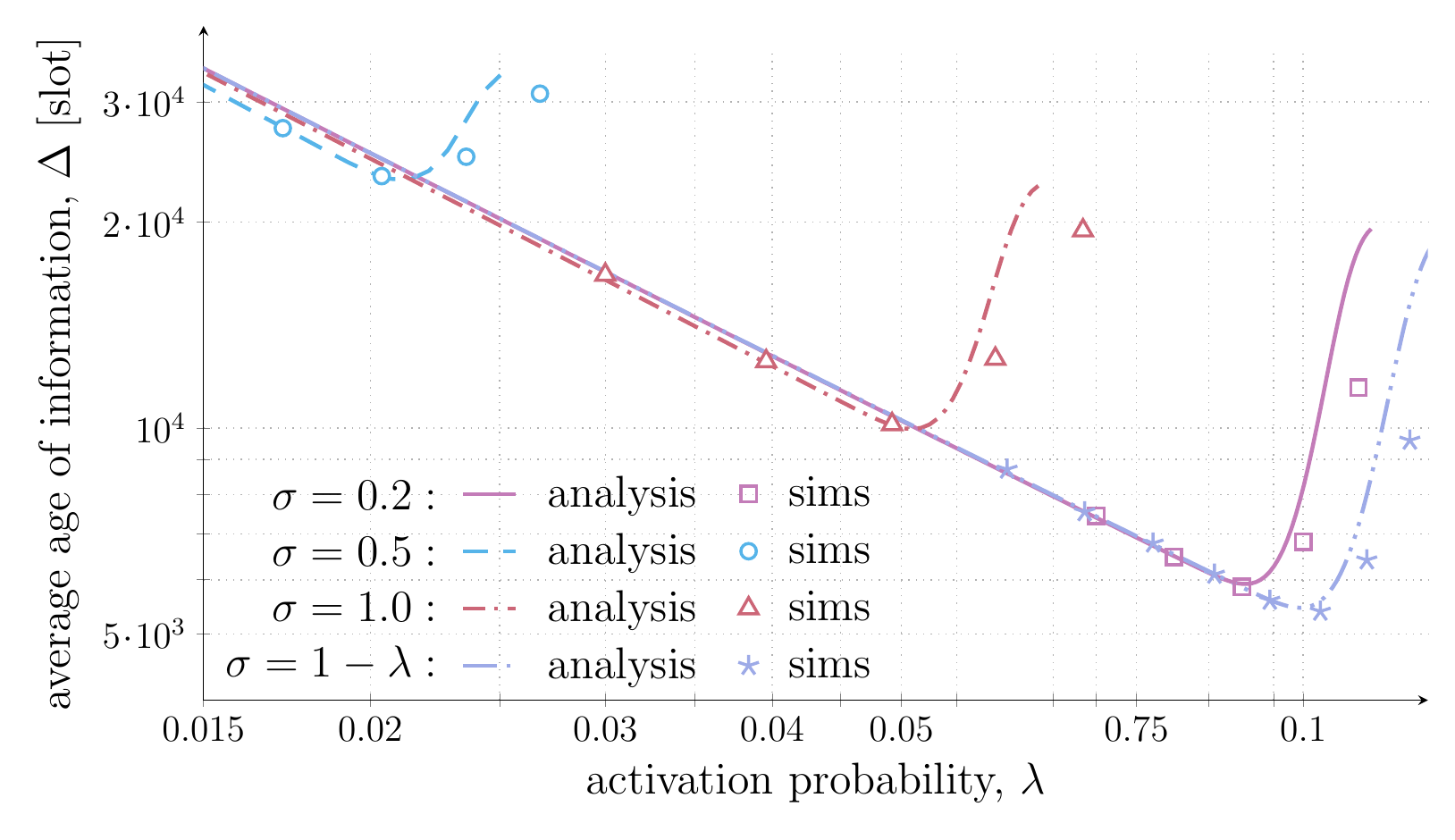}
  \caption{Average AoI for IRSA vs probability $\lambda$ of moving from silent to active state. Different lines denote the behaviour for different probabilities of moving from active to silent state (\pas). In all cases, $\Lambda(x)=x^3$ and $\nodes=4000$.}
  \vspace{-.5em}
  \label{fig:aoiBursty}
\end{figure}

This result can conveniently be used within the analytical framework developed for IRSA in Sec.~\ref{sec:analysis}.  Let us focus, for instance, on the average AoI. As shown in the proof of Prop.~\ref{prop:avgAoI}, $\avAoI$ can be directly computed once the stationary distribution of the current age at the start of a frame is available. Specifically, if we employ \eqref{eq:avgNodeAgeStationary} and recall that in our case $\apRV_k$ takes values in the set $\{\slots\ell\}$, $\ell\geq0$, we immediately obtain
\begin{align}
  \avAoI = 1 + \frac{3\slots}{2} + \sum_{\ell=0}^\infty \slots\,\ell \cdot \pi_{\ell}
\label{eq:avgAgeBursty}
\end{align}
which, fed with the solution of \eqref{eq:matrixSolutionBursty}, provides the average AoI for the traffic model under study. The formulation confirms the broad applicability of the proposed Markovian approach.

Leaning on this, Fig.~\ref{fig:aoiBursty} explores the impact of different activation patterns on the average AoI. On the $x$-axis we report the probability $\psa$, which is a direct indicator of how often a terminal initiates a transmission burst. Note in fact that, after entering the silent state, a node will become active again after $1/\psa$ frames on average. Different lines indicate the behaviour in terms of average AoI for distinct values of $\pas$. This reflects the duration of a burst, as an active node will transmit on average new updates for $1/\sigma$ frames in a row before returning silent. From this standpoint, two curves are of special relevance. First, the case $\sigma=1$, which corresponds to having a node immediately going back to silent mode after the transmission of an update. Second, the case $\sigma=1-\lambda$ (solid line), which allows a direct comparison to the traffic pattern studied in Sec.~\ref{sec:analysis}. As discussed, in fact, this corresponds to having terminal become active at each frame with i.i.d. probability $\psa$.

Besides illustrating the AoI performance for a variety of traffic configurations, Fig.~\ref{fig:aoiBursty} provides some interesting insights. When nodes experience long periods of silence (low \psa), transmitting a sequence of updates (low \pas) may be beneficial in terms of AoI. This approach, indeed, boosts the probability that at least one of the packets reaches the destination. Conversely, when \psa\ increases, a bursty behaviour becomes detrimental. This is due to the higher channel load induced by the larger number of transmissions, eventually leading to higher packet loss rates.  In this perspective, it is particularly interesting to note how the lowest average AoI is in fact obtained for $\pas=1$, which outperform event the i.i.d. traffic case (solid line). Such a result offers a non-trivial design hint, suggesting how the drop of some generated updates may be beneficial in terms of AoI under congested channel conditions.

\section{Conclusions}
\label{sec:conclusions}

This paper offered the first analytical characterisation of a class of modern random access scheme from an information freshness standpoint. Specifically, focusing on the irregular repetition slotted ALOHA (IRSA) protocol, the evolution of the current \ac{AoI} for a status update source has been tracked by means of a Markov chain, which was proven to be ergodic. Leaning on a closed-form expression for the stationary distribution of such process, simple exact expressions for the average \ac{AoI} and for the age violation probability were discussed. The presented analysis revealed how IRSA can significantly outperform slotted ALOHA with respect to both metrics, and pinpointed some fundamental trade-offs that drive the behaviour of the protocol, capturing the impact of frame length and replica number distribution. The compact formulations introduced in the article offer a simple yet effective tool for proper system design, as illustrated in the numerical example provided in App.~\ref{app:appendixLoRa}.

\begin{appendices}
\section{} \label{app:appendixLoRa}

To illustrate how the derived framework can serve as a useful system design tool, consider the following example. Let us focus on an IoT setup where devices monitor a physical quantity, and report the measurements to a central unit over a shared channel. Transmissions are performed at a bitrate of $3.125$ kbps, and the transmission of a packet takes $T=136$ ms (duration of a slot in our case). Such parameters are compatible with the LoRaWAN standard, considering a payload of $32$ bytes, explicit header mode and a bandwidth of $125$ kHz \cite{Zanella20_IoT}. Assume furthermore that each device performs a new reading \--- and transmits a status update \--- on average every $A=10$ minutes. In this configuration, we are interested in evaluating how many devices can be supported so that the average AoI does not exceed the target value $\Delta^\prime = 10.5$ minutes. Note that this is a particularly demanding design. Indeed, we aim to keep an average AoI for each node which is very close to its average update generation rate (an ultimate lower bound), and want to do so having all devices share the channel in a fully uncoordinated fashion. For \ac{SA}, the design aspect can be targeted by solving \eqref{eq:resultsSA} with respect to \nodes, to obtain
\begin{align*}
\nodes_{\mathsf{sa}} \leq \left\lfloor 1 - \frac{\ln\big(\pAct(\Delta^\prime/T - 1/2)\big)}{\ln(1-\pAct)} \right\rfloor = 215 \,\textrm{ [devices]}
\end{align*}
where $\Delta^\prime$ has been normalised to the slot duration, and $\pAct = T/A$ is the activation probability per slot corresponding to the average generation rate $A$. Let us now focus on IRSA. First of all, we observe that the frame-based operations pose a constraint to the minimum average AoI that can be achieved, since longer frames also entail higher latency for delivering an update. The maximum frame size that may be used to support $\Delta^\prime$ can immediately be derived from \eqref{eq:avgAoIIRSA} and \eqref{eq:pGeoIRSA}, considering the limiting condition of having a single device ($\nodes=1$) and assuming no packet loss. In this situation, the average AoI in \eqref{eq:avgAoIIRSA} evaluates to $\Delta = 3\slots/2 + 1/\pAct$ slots. Solving with respect to \slots, we obtain that the IRSA shall be operated with frames satisfying
\begin{align*}
\slots \leq \left\lfloor \frac{2}{3} \left(\Delta^\prime/T - 1/\pAct \right) \right \rfloor = 147 \, \text{ [slots]}.
\end{align*}
Consider then a frame of size $\slots = 100$ slots. Equation \eqref{eq:avgAoIIRSA} can then easily be solved numerically with respect to \nodes\ to obtain the maximum number of nodes that can be supported without exceeding the average age constraint, to obtain $\nodes_{\mathsf{irsa}} \leq 2600$ [devices], with a more than ten-fold improvement over the slotted ALOHA solution.

\section{} \label{app:appendixIRSA}

For completeness, we report the IRSA packet loss rate approximations employed in \eqref{eq:plr_approx} for the degree distribution $\Lambda(x)=x^3$. In the error-floor region, we have \cite{Sandgren17:FACSA}:
\begin{align}
  \plossef \simeq \sum_{s=1}^{2} \frac{\phi(s) \nu(s) c(s)}{\nu(s)!} \binom{\slots}{\mu(s)} \binom{\slots}{3}^{-\nu(s)}
  \label{eq:apprErrFloor}
  \end{align}
  where $x(s)$ denotes the $s$-th element of the generic vector $x$, and the vectors used in the equation are given as $\nu = [2, \,3]$, $\mu = [3,\, 4]$, $c = [1,\, 24]$ and
  \begin{align*}
    \phi(s) = \sum_{k=0}^{\nu(s)-1} (-1)^{\nu(s)-1+k} (\slots \load)^k \frac{(\nu(s)-1)!}{k!}, \quad s = 1,2.
  \end{align*}
In the waterfall region we employ instead  \cite{Graell18_Waterfall}
\begin{align}
  \plosswf \simeq \gamma Q\left( \frac{\sqrt{\slots} ( \load^* - \beta_0 \slots^{-2/3} - \load )}{\sqrt{\alpha_0^2 + \load(1-\slots\load/\nodes)}} \right)
\end{align}
where $\gamma=0.784399$, $\load^*=0.818469$, $\alpha_0=0.497867$, $\beta_0=0.964528$. For further details as well as for the expressions under different degree distributions, we refer to \cite{Sandgren17:FACSA} and \cite{Graell18_Waterfall}.

\section*{Acknowledgements}
The Author would like to thank Gianluigi Liva and Giuseppe Durisi for the insightful discussions on the topic.

\end{appendices}

\bibliographystyle{IEEEtran}
\bibliography{IEEEabrv,aloha}

\begin{thebibliography}{10}
\providecommand{\url}[1]{#1}
\csname url@samestyle\endcsname
\providecommand{\newblock}{\relax}
\providecommand{\bibinfo}[2]{#2}
\providecommand{\BIBentrySTDinterwordspacing}{\spaceskip=0pt\relax}
\providecommand{\BIBentryALTinterwordstretchfactor}{4}
\providecommand{\BIBentryALTinterwordspacing}{\spaceskip=\fontdimen2\font plus
\BIBentryALTinterwordstretchfactor\fontdimen3\font minus
  \fontdimen4\font\relax}
\providecommand{\BIBforeignlanguage}[2]{{%
\expandafter\ifx\csname l@#1\endcsname\relax
\typeout{** WARNING: IEEEtran.bst: No hyphenation pattern has been}%
\typeout{** loaded for the language `#1'. Using the pattern for}%
\typeout{** the default language instead.}%
\else
\language=\csname l@#1\endcsname
\fi
#2}}
\providecommand{\BIBdecl}{\relax}
\BIBdecl

\bibitem{Liva11:IRSA}
G.~Liva, ``Graph-based analysis and optimization of contention resolution
  diversity slotted {ALOHA},'' \emph{{IEEE} Trans. Commun.}, vol.~59, no.~2,
  pp. 477--487, 2011.

\bibitem{Saad20_6G}
W.~{Saad}, M.~{Bennis}, and M.~{Chen}, ``A vision of {6G} wireless systems:
  Applications, trends, technologies, and open research problems,'' \emph{IEEE
  Network}, vol.~34, no.~3, pp. 134--142, 2020.

\bibitem{Munari20_6G}
\BIBentryALTinterwordspacing
{6G Flagship}, ``Critical and massive machine type communication towards
  {6G},'' 2020. [Online]. Available: \url{http://arxiv.org/abs/2004.14146}
\BIBentrySTDinterwordspacing

\bibitem{Popovski19_Wiley}
I.~Leyva‐Mayorga, C.~Stefanovic, P.~Popovski, V.~Pla, and
  J.~Martinez‐Bauset, ``Random access for machine‐type communications,''
  \emph{Wiley 5G Ref}, 2019.

\bibitem{Abramson:ALOHA}
N.~Abramson, ``The {ALOHA} system - another alternative for computer
  communications,'' in \emph{Proc. 1970 Fall Joint Computer Conference}.\hskip
  1em plus 0.5em minus 0.4em\relax AFIPS Press, 1970.

\bibitem{LoRa}
{LoRa Alliance Technical Commitee}, ``{LoRaWAN 1.1 Specification}.''

\bibitem{SigFox}
Sigfox, ``{SIGFOX: The Global Communications Service Provider for the Internet
  of Things},'' \url{www.sigfox.com}.

\bibitem{Wang17_CommMag}
Y.~{Wang}, X.~{Lin}, A.~{Adhikary}, A.~{Grovlen}, Y.~{Sui}, Y.~{Blankenship},
  J.~{Bergman}, and H.~{Razaghi}, ``A primer on {3GPP} narrowband internet of
  things,'' \emph{IEEE Commun. Mag.}, vol.~55, no.~3, pp. 117--123, 2017.

\bibitem{Berioli2016}
M.~Berioli, G.~Cocco, G.~Liva, and A.~Munari, \emph{{Modern random access
  protocols}}.\hskip 1em plus 0.5em minus 0.4em\relax NOW Publisher, 2016.

\bibitem{Paolini15:TIT_CSA}
E.~Paolini, G.~Liva, and M.~Chiani, ``Coded slotted {ALOHA}: A graph-based
  method for uncoordinated multiple access,'' \emph{{IEEE} Trans. Inf. Theory},
  vol.~61, no.~12, pp. 6815--6832, 2015.

\bibitem{Clazzer18:ECRA}
F.~Clazzer, C.~Kissling, and M.~Marchese, ``Enhancing contention resolution
  {ALOHA} using combining techniques,'' \emph{{IEEE} Trans. Commun.}, vol.~66,
  no.~6, pp. 2576--2587, 2018.

\bibitem{Polyanskiy17:RA}
Y.~Polyanskiy, ``A perspective on massive random-access,'' in \emph{Proc. IEEE
  ISIT}, 2017.

\bibitem{Stefanovic13:RatelessAloha}
C.~Stefanovic and P.~Popovski, ``{ALOHA} random access that operates as a
  rateless code,'' \emph{{IEEE} Commun. Mag.}, vol.~61, no.~11, pp. 4653--4662,
  2013.

\bibitem{Chamberland20_TIT}
V.~K. {Amalladinne}, J.~F. {Chamberland}, and K.~R. {Narayanan}, ``A coded
  compressed sensing scheme for unsourced multiple access,'' \emph{{IEEE}
  Trans. Inf. Theory}, vol.~66, no.~10, pp. 6509--6533, 2020.

\bibitem{Caire19:ISIT}
A.~Fengler, P.~Jung, and G.~Caire, ``{SPARC}s and {AMP} for unsourced random
  access,'' \emph{in Proc. IEEE ISIT}, 2019.

\bibitem{Frolov20_TCOM}
S.~{Kowshik}, K.~{Andreev}, A.~{Frolov}, and Y.~{Polyanskiy}, ``Energy
  efficient coded random access for the wireless uplink,'' \emph{{IEEE} Trans.
  Commun.}, vol.~68, no.~8, pp. 4694--4708, 2020.

\bibitem{dvbrcs2}
{ETSI}, ``{EN 301 545-2}: Digital video broadcasting {(DVB)}; second generation
  {DVB} interactive satellite system {(DVB-RCS2)}; part 2: Lower layers for
  satellite standard,'' Tech. Rep.

\bibitem{Uysal20_TIT}
Y.~Sun, Y.~Polyanskiy, and E.~Uysal, ``Sampling of the {Wiener} process for
  remote estimation over a channel with random delay,'' \emph{{IEEE} Trans.
  Inf. Theory}, vol.~66, no.~2, pp. 1118--1135, 2020.

\bibitem{Ephremides19_AoII}
\BIBentryALTinterwordspacing
A.~Maatouk, S.~Kriouile, M.~Assaad, and A.~Ephremides, ``The age of incorrect
  information: a new performance metric for status updates,'' 2019. [Online].
  Available: \url{http://arxiv.org/abs/1907.06604v1}
\BIBentrySTDinterwordspacing

\bibitem{Soleymani20_valueInfo}
T.~Soleymani, J.~Baras, and S.~Hirche, ``Value of information in feedback
  control,'' \emph{{IEEE} Trans. Autom. Control}, 2020.

\bibitem{Kaul11_SECON}
S.~{Kaul}, M.~{Gruteser}, V.~{Rai}, and J.~{Kenney}, ``Minimizing age of
  information in vehicular networks,'' in \emph{Proc. IEEE SECON}, June 2011.

\bibitem{Kaul11_Globecom}
S.~{Kaul}, R.~{Yates}, and M.~{Gruteser}, ``On piggybacking in vehicular
  networks,'' in \emph{Proc. IEEE GLOBECOM}, Dec 2011.

\bibitem{Yates19_TIT}
R.~D. {Yates} and S.~K. {Kaul}, ``The age of information: Real-time status
  updating by multiple sources,'' \emph{{IEEE} Trans. Inf. Theory}, vol.~65,
  no.~3, pp. 1807--1827, March 2019.

\bibitem{Modiano19_book}
Y.~{Sun}, I.~{Kadota}, R.~{Talak}, E.~{Modiano}, and R.~{Srikant}, \emph{Age of
  Information: A New Metric for Information Freshness}.\hskip 1em plus 0.5em
  minus 0.4em\relax Morgan \& Claypool, 2019.

\bibitem{Zhou20_TWC}
B.~{Zhou} and W.~{Saad}, ``Minimum age of information in the internet of things
  with non-uniform status packet sizes,'' \emph{{IEEE} Trans. Wireless
  Commun.}, vol.~19, no.~3, pp. 1933--1947, 2020.

\bibitem{Bedewy19}
A.~Bedewy, Y.~Sun, S.~Kompella, and N.~Shroff, ``Age-optimal sampling and
  transmission scheduling in multi-source systems,'' in \emph{Proc. ACM
  MOBIHOC}, 2019.

\bibitem{Zhou19_TCOM}
B.~{Zhou} and W.~{Saad}, ``Joint status sampling and updating for minimizing
  age of information in the internet of things,'' \emph{{IEEE} Trans. Commun.},
  vol.~67, no.~11, pp. 7468--7482, 2019.

\bibitem{Jiang19_IoT}
Z.~{Jiang}, B.~{Krishnamachari}, X.~{Zheng}, S.~{Zhou}, and Z.~{Niu}, ``Timely
  status update in wireless uplinks: Analytical solutions with asymptotic
  optimality,'' \emph{IEEE Internet of Things Journal}, vol.~6, no.~2, pp.
  3885--3898, 2019.

\bibitem{Modiano19_TNET}
I.~{Kadota}, A.~{Sinha}, and E.~{Modiano}, ``Scheduling algorithms for
  optimizing age of information in wireless networks with throughput
  constraints,'' \emph{IEEE/ACM Trans. Netw.}, vol.~27, no.~4, pp. 1359--1372,
  2019.

\bibitem{Ephremides19_Infocom}
A.~{Maatouk}, M.~{Assaad}, and A.~{Ephremides}, ``Minimizing the age of
  information: {NOMA} or {OMA}?'' in \emph{Proc. IEEE INFOCOM Workshops}, April
  2019.

\bibitem{Yates17:AoI_SA}
R.~Yates and S.~Kaul, ``Status updates over unreliable multiaccess channels,''
  in \emph{Proc. IEEE ISIT}, June 2017.

\bibitem{Modiano18_AoI}
R.~Talak, S.~Karaman, and E.~Modiano, ``Distributed scheduling algorithms for
  optimizing information freshness in wireless networks,'' in \emph{Proc. IEEE
  SPAWC}, June 2018.

\bibitem{Shirin19_ISIT}
X.~Chen, K.~Gatsis, H.~Hassani, and S.~Bidokhti, ``Age of information in random
  access channels,'' in \emph{Proc. IEEE ISIT}, 2020.

\bibitem{Yates20_ISIT}
R.~Yates and S.~Kaul, ``Age of information in uncoordinated unslotted
  updating,'' in \emph{{Proc. IEEE ISIT}}, 2020.

\bibitem{Munari20_avgAoI}
A.~Munari and A.~Frolov, ``Average age of information of irregular repetition
  slotted {ALOHA},'' in \emph{{Proc. IEEE Globecom}}, 2020.

\bibitem{Sandgren17:FACSA}
E.~Sandgren, A.~Graell~i Amat, and F.~Br\"annstr\"om, ``On frame asynchronous
  coded slotted {ALOHA}: Asymptotic, finite length, and delay analysis,''
  \emph{{IEEE} Trans. Commun.}, vol.~65, no.~2, pp. 691--703, 2017.

\bibitem{Graell18_Waterfall}
A.~{Graell i Amat} and G.~{Liva}, ``Finite-length analysis of irregular
  repetition slotted {ALOHA} in the waterfall region,'' \emph{IEEE Commun.
  Letters}, vol.~22, no.~5, pp. 886--889, 2018.

\bibitem{GallagerStochasticProc}
R.~Gallager, \emph{{Stochastic Processes: Theory for Applications}}.\hskip 1em
  plus 0.5em minus 0.4em\relax Cambridge University Press, 201.

\bibitem{Ephremides14_peakAge}
M.~{Costa}, M.~{Codreanu}, and A.~{Ephremides}, ``Age of information with
  packet management,'' in \emph{Proc. IEEE ISIT}, 2014.

\bibitem{Durisi19_JSAC}
R.~{Devassy}, G.~{Durisi}, G.~C. {Ferrante}, O.~{Simeone}, and E.~{Uysal},
  ``Reliable transmission of short packets through queues and noisy channels
  under latency and peak-age violation guarantees,'' \emph{{IEEE} J. Sel. Areas
  Commun.}, vol.~37, no.~4, pp. 721--734, April 2019.

\bibitem{Popovski20_CommLett}
F.~Chiariotti, O.~Vikhrova, B.~Soret, and P.~Popovski, ``Peak age of
  information distribution in tandem queue systems,'' \emph{{IEEE Comm.
  Letters}}, 2020.

\bibitem{Frolov19_tFoldSA}
A.~{Glebov}, N.~{Matveev}, K.~{Andreev}, A.~{Frolov}, and A.~{Turlikov},
  ``Achievability bounds for {T-fold} irregular repetition slotted {ALOHA}
  scheme in the {Gaussian MAC},'' in \emph{Proc. IEEE WCNC}, 2019.

\bibitem{Zanella20_IoT}
D.~{Magrin}, M.~{Capuzzo}, and A.~{Zanella}, ``A thorough study of {LoRaWAN}
  performance under different parameter settings,'' \emph{IEEE Internet of
  Things Journal}, vol.~7, no.~1, pp. 116--127, 2020.

\end{thebibliography}


\end{document}